\documentclass[journal]{IEEEtran}

\usepackage{cite}
\usepackage{amsmath,amssymb,amsfonts,mathtools,bm,dsfont}
\usepackage{graphicx}
\usepackage[caption=false,font=footnotesize]{subfig}
\usepackage{booktabs}
\usepackage{xcolor}
\usepackage{algorithm}
\usepackage{algpseudocode}
\usepackage[most]{tcolorbox}
\usepackage{enumitem}
\usepackage{balance}
\usepackage{tikz}
\usepackage[normalem]{ulem}
\usepackage{hyperref}

\usetikzlibrary{positioning,shapes.geometric,arrows}

\hypersetup{
  colorlinks=true,
  linkcolor=blue,
  citecolor=blue,
  urlcolor=blue
}

\begin{document}

\title{Verifiable Manifest Signing and Transparency Enforcement for Secure MCP-Based LLM Pipelines}

\author{
Saeid~Jamshidi,
Kawser~Wazed~Nafi,
Arghavan~Moradi~Dakhel,
Foutse~Khomh,
and Mohammad~Hamdaqa%
\thanks{
S. Jamshidi, K. W. Nafi, A. Moradi Dakhel, and F. Khomh are with the SWAT Laboratory, Polytechnique Montréal, Montréal, Canada.
}
\thanks{
M. A. Hamdaqa is with the SæT Laboratory, Polytechnique Montréal, Montréal, Canada.
}
}

\maketitle

\begin{abstract}
Large Language Models (LLMs) are increasingly integrated into tool-driven environments, e.g., healthcare analytics, financial systems, retrieval-augmented generation, and multi-agent workflows. Although the Model Context Protocol (MCP) standardizes how LLM applications expose and invoke external tools, its baseline execution model does not require tool-use manifests to be cryptographically authenticated, freshness-checked, policy-bound, and independently auditable before execution. As a result, MCP-based pipelines can remain vulnerable to manifest tampering, unauthorized tool invocation, replay of stale execution requests, and weak post-execution accountability. This paper introduces a manifest-level enforcement layer for MCP-based LLM pipelines. The key contribution is to treat each MCP tool-use manifest as a first-class security object whose canonical representation must be policy-validated, freshness-checked, digitally signed, verified before execution, and linked to tamper-evident audit evidence. The proposed approach specifically binds MCP tool invocation to verifiable manifest integrity and fail-closed execution authorization. The system separates user-visible request parameters from model-side execution metadata, rejects non-compliant and stale manifests before execution, and records accepted invocations in a Merkle-based transparency log. Experimental evaluation across GPT-5.3, LLaMA-3.5, and DeepSeek-V3 using workloads up to 50{,}000 manifest instances demonstrates near-linear scalability ($R^2 = 0.998$), bounded verification latency ($\le 9.4$ ms on edge devices), and reliable rejection of expired, malformed, replayed, and policy-violating manifests (rejection rate $>98.7\%$). Additional MCP-oriented experiments across healthcare, financial, RAG, and multi-agent scenarios show that manifest-level cryptographic enforcement can provide low-overhead, traceable, and auditable execution control for heterogeneous LLM-tool pipelines.
\end{abstract}

\section{Introduction}
\label{Intro}
Large Language Models (LLMs) are increasingly integrated into tool-driven environments, including healthcare analytics, financial systems, and autonomous decision-support platforms~\cite{openai2023gpt4,touvron2023llama,deepseek2024llm}. In these settings, LLMs no longer operate as isolated conversational systems; instead, they interact with external tools, APIs, databases, retrieval engines, and execution services \cite{wu2024isolategpt, reddy2025modular}. Consequently, the reliability and security of the execution pipeline become as critical as the model's reasoning capability. A correct model response may still lead to unsafe behavior if the associated tool invocation is unauthorized, unverifiable, stale, and executed without sufficient auditability \cite{tanveer2025towards, anbiaee2026security}.\\
For example, consider an LLM-assisted healthcare analytics system in which a user asks the model to summarize patient-level laboratory results and query a clinical database. The LLM may need to invoke an external data-access tool with parameters, e.g., the requested record type, access scope, timestamp, and execution context \cite{ray2025survey, sonkar2025llm}. Without a verifiable execution layer, an attacker could manipulate the tool request, replay an outdated manifest, alter hidden execution metadata, and trigger an unauthorized API call while the user-visible response appears normal \cite{liu2026dive}. Similarly, in a financial analytics pipeline, a manipulated tool invocation could query sensitive records, execute an unintended transaction-related function, and bypass policy restrictions. These examples illustrate that tool-integrated LLM security depends not only on model alignment, but also on verifiable control over the execution artifacts that mediate model-tool interactions \cite{meng2026agent, chu2026stateless}. The Model Context Protocol (MCP) has recently emerged as a standardized approach for structuring interactions between LLMs and external tools~\cite{mcp-spec,anthropic2024mcp,xin2025mcpguard}. MCP defines interaction semantics, manifest structures, and policy interfaces that support controlled communication between models and execution environments \cite{venkiteela2025new}. In particular, MCP encourages separating model-internal execution metadata from user-visible content, which can reduce unintended information exposure and improve execution consistency. However, baseline MCP specifications do not inherently provide cryptographic enforcement, transparent runtime verification, and tamper-evident accountability guarantees for executed actions \cite{errico2025securing, hou2025model}. Therefore, while MCP improves the organization of model-tool interactions, additional enforcement mechanisms are required to verify that tool-use manifests are authentic, policy-compliant, up to date, and auditable before execution \cite{ray2025survey}. These limitations become increasingly critical in safety-sensitive environments (e.g., finance, healthcare, and public governance), where even minor execution inconsistencies and unauthorized tool invocations can lead to outages, compliance violations, privacy breaches, and Denial-of-Service (DoS) risks. Existing LLM deployment pipelines remain vulnerable to several execution-layer threats, including unverifiable tool invocation, adversarial prompt manipulation, asymmetric key concentration, replay of outdated execution artifacts, and opaque runtime behavior~\cite{liu2025exploit,zhang2025large}. For instance, Tool Invocation Prompt (TIP) attacks can manipulate execution flows via malicious prompt injection~\cite{liu2025exploit}, while timing irregularities may reveal workload patterns via side-channel leakage. Similarly, excessive dependence on a limited subset of signing keys can amplify the operational impact of credential compromise~\cite{hou2025model}.\\
Despite rapid progress in model alignment, robustness, and reasoning performance, most existing studies remain largely model-centric~\cite{zhang2025large}. Comparatively less attention has been given to execution-layer properties, e.g., manifest integrity enforcement, runtime verification, auditability, and cryptographic accountability. Existing approaches, including runtime attestation~\cite{su2025runtime}, enclave-based verification systems (e.g., Scanclave~\cite{DBLP:journals/corr/abs-1907-09906}), and automated rule-generation methods~\cite{greco2025formal}, improve specific aspects of trust and compliance. However, they do not provide a unified enforcement layer for probabilistic, tool-integrated LLM pipelines that combine manifest validation, digital signing, transparency logging, and audit export. Consequently, current LLM ecosystems still face challenges in tampering detection, runtime traceability, consistent policy enforcement, and operational transparency \cite{bollikonda2025secure}.\\
To address these challenges, this paper presents a manifest-level enforcement approach for MCP-based LLM tool pipelines. Instead of proposing a broad security framework, the approach focuses on one execution artifact: the MCP tool-use manifest. The central idea is to treat each manifest as a first-class security object that must be canonically encoded, policy-validated, freshness-checked, digitally signed, verified before execution, and linked to tamper-evident audit evidence. In this design, each tool invocation is represented by a structured manifest containing user-visible request fields, model-side execution metadata, and freshness information. Only manifests that satisfy policy and freshness constraints are signed using protected signing keys and admitted to execution. Accepted manifests are then recorded in an append-only transparency log, enabling system operators and external auditors to verify whether a tool invocation was authorized, fresh, intact, and traceable. The proposed approach is implemented and evaluated across three representative LLM backends, GPT 5.3, LLaMA-3.5, and DeepSeek-V3, using workloads of up to 50{,}000 manifest instances. Experimental findings demonstrate near-linear scalability ($R^2=0.998$), stable verification behavior as workload size increases, effective rejection of malformed and policy-violating manifests, and balanced utilization across the evaluated LLMs. The results also reveal deployment-level considerations, e.g., key allocation imbalance, which motivate adaptive key management and rotation strategies in future MCP-based execution systems.
\begin{itemize}
    \item We introduce a manifest-level enforcement approach for MCP-based LLM tool pipelines that treats each tool-use manifest as a first-class security object before execution authorization.
    
    \item We design an MCP-specific execution workflow that binds each manifest to canonical encoding, policy validation, freshness checking, digital signing, pre-execution verification, and tamper-evident audit evidence.
    
    \item We develop a reproducible evaluation methodology that combines scalability, verification reliability, utilization balance, and operational security metrics to assess manifest-level enforcement under large-scale workloads.
    
    \item We provide empirical insights into MCP execution-layer behavior, including verification stability, timing variance, LLM utilization balance, workload-dependent enforcement dynamics, and signing key concentration effects.
\end{itemize}

The paper is organized as follows: Section~\ref{RelatedWork} reviews prior work and gaps, Section~\ref{sec:RQs} presents research questions, Section~\ref{Proposed} details the solution and threat model, Section~\ref{Evaluation} outlines the experimental setup, Section~\ref{sec:Findings} reports results, Sections~\ref{sec:discussion},~\ref{sec:validity}, and~\ref{sec:limitations} discuss implications, validity, and limitations, and Section~\ref{sec:conclusion} concludes with future directions.

\section{Related Work}
\label{RelatedWork}
Research on trustworthy LLM deployment spans transparency, runtime attestation, adversarial tool-invocation security, and compliance-aware orchestration.

\subsection{Transparency and Auditability}
Transparency and accountability mechanisms have been extensively studied in distributed and security-critical systems. Hicks~\cite{hicks2022sok} systematizes transparency technologies based on logging, sanitization, and query verification. While these approaches establish important foundations for accountability, they were primarily designed for deterministic infrastructures, e.g., certificate transparency systems, rather than for probabilistic, tool-driven LLM environments. Similarly, Reijsbergen et al.~\cite{reijsbergen2023tap} introduce TAP, a transparency-preserving architecture based on authenticated data structures and zero-knowledge proofs. Although TAP improves verifiability in structured multi-user systems, its applicability to dynamic LLM execution and tool orchestration remains limited.

\subsection{Runtime Attestation and Secure Execution}
Several studies investigate runtime integrity and attestation in distributed environments. Su et al.~\cite{su2025runtime} propose continuous verification mechanisms for cloud workloads, while Scanclave~\cite{DBLP:journals/corr/abs-1907-09906} leverages enclave-based protection for secure execution. These approaches improve infrastructure-level trust and execution integrity; however, they do not explicitly address LLM-specific execution semantics, e.g., adversarial prompts, chained tool interactions, probabilistic runtime behavior, and manifest-level policy enforcement. In addition, enclave-dependent architectures may introduce scalability and deployment constraints in heterogeneous LLM ecosystems.

\subsection{Adversarial Tool Invocation and Prompt Security}
Other studies primarily focus on identifying vulnerabilities in LLM pipelines. Liu et al.~\cite{liu2025exploit} demonstrate that adversarial prompts can manipulate tool invocation behavior, exposing the fragility of unverified execution pipelines. Their findings highlight the importance of enforcing validation and policy checks before runtime. However, these approaches mainly characterize threats and attacks rather than providing end-to-end cryptographic enforcement and transparent verification mechanisms.

\subsection{Compliance and LLM Governance}
Compliance-aware orchestration and governance have also received growing attention. Greco et al.~\cite{greco2025formal} explore automated rule generation and policy synthesis using LLMs. While such approaches improve regulatory automation and policy generation, they provide limited guarantees regarding runtime enforcement, manifest validation, and execution traceability. More broadly, surveys and technical reports on LLM safety and deployment~\cite{zhang2025large,openai2023gpt4,touvron2023llama,deepseek2024llm,deepseek2025survey} extensively discuss robustness, fairness, alignment, and efficiency. Nevertheless, comparatively limited attention has been given to execution-layer properties, e.g., cryptographic accountability, transparent auditability, and secure orchestration of external tools.\\

The literature synthesis indicates that existing studies primarily address isolated aspects of trustworthy LLM deployment, including transparency, runtime attestation, adversarial analysis, and policy generation. However, current approaches rarely integrate cryptographic manifest validation, runtime verification, transparency-aware logging, and scalable execution monitoring within a single execution-layer solution. To address this gap, this work introduces a secure tool manifest and digital signing solution that extends MCP with cryptographically verifiable manifests, tamper-evident audit logging, and runtime validation mechanisms for tool-integrated LLM pipelines.

\section{Proposed Methodology}
\label{Proposed}
\begin{figure*}[h]
    \centering
    \includegraphics[width=0.80\textwidth]{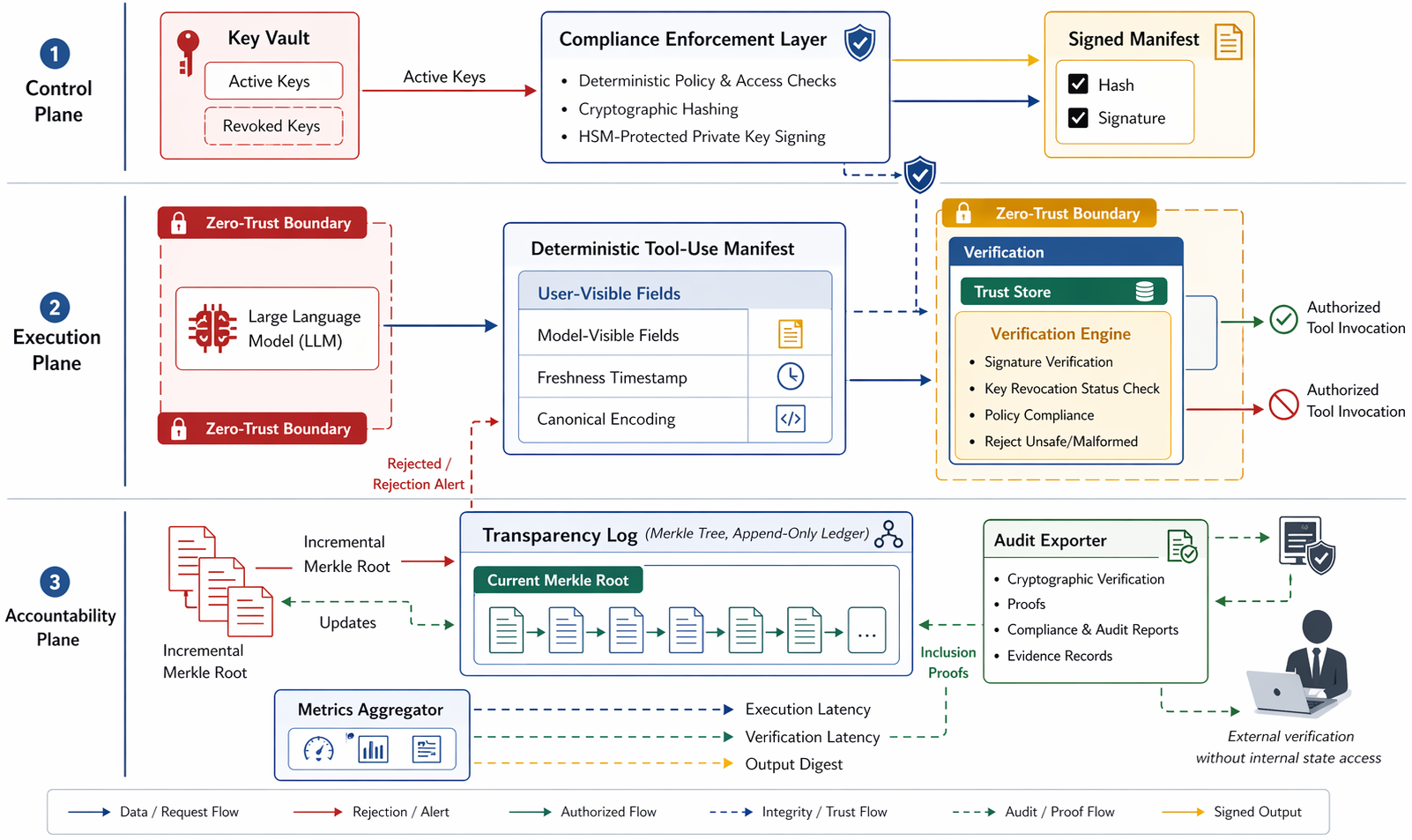}
    \caption{End-to-end architecture of the proposed secure manifest-enforcement solution, illustrating manifest creation, policy enforcement, signing, verification, transparency logging, audit export, and metrics collection across the MCP-based LLM execution lifecycle.}
    \label{fig:architecture}
\end{figure*}

This section presents the proposed manifest-level enforcement approach for MCP-based, tool-integrated LLM pipelines. The approach focuses on a single execution artifact: the MCP tool-use manifest. Although digital signatures, runtime verification, transparency logs, and audit records are established security primitives, their role in this work is to enforce the integrity and authorization status of MCP manifests before tool execution. Thus, the contribution is not a new cryptographic primitive, but an MCP-specific enforcement workflow that binds each tool invocation to a canonical, policy-compliant, freshness-valid, and verifiable manifest. Rather than modifying the core MCP interaction model, the proposed approach strengthens the execution layer by requiring every tool request to pass through manifest-centered authorization. As illustrated in Figure~\ref{fig:architecture}, the workflow consists of six sequential stages: 1) manifest creation, 2) policy enforcement and signing, 3) verification, 4) transparency logging, 5) audit exporting, and 6) metrics collection. The execution flow begins when a tool request is converted into a structured manifest containing user-visible fields, model-side execution metadata, and freshness information. The manifest is then checked against policy rules and freshness constraints. Only compliant manifests are digitally signed and admitted to the execution path. Before execution, the signed manifest is verified; after acceptance, it is appended to a tamper-evident transparency log and exported as compact audit evidence. Runtime metrics are collected throughout the process to evaluate latency, verification overhead, and operational stability. This design reduces execution-layer risks associated with unauthorized tool invocation, replayed and malformed manifests, key misuse, timing irregularities, and manipulation of tool-execution metadata. More importantly, it makes MCP tool execution externally verifiable: system operators and auditors can verify whether a tool invocation was authorized, fresh, intact, and traceable without relying solely on LLM output and opaque runtime behavior.

\subsection{Threat Model}
The threat model defines the adversarial capabilities, trust assumptions, and security boundaries that the proposed solution considers. The focus is on execution-layer threats in tool-integrated LLM pipelines, including manifest tampering, unauthorized tool invocation, replay attacks, freshness violations, verification failures, and audit manipulation. The objective is not to secure the LLM's internal reasoning process, but to ensure that tool execution occurs only through policy-compliant, signed, verifiable, and transparently logged manifests.
We assume a probabilistic polynomial-time (PPT) adversary $\mathcal{A}$~\cite{klooss2021expected} capable of observing, delaying, replaying, intercepting, and modifying communications between the layer, tool-execution layer, verification service, and audit components. The adversary may inject malformed manifests, alter manifest fields, submit expired manifests, attempt to access unauthorized tools, and reuse previously valid manifests outside their freshness window. These capabilities represent practical execution-layer attacks against manifest integrity and policy enforcement. The proposed solution assumes that collision-resistant hash functions, digital signature primitives, and HSM-protected key management components \cite{talsania2026robust} remain trusted. In particular, the private signing key is assumed to be inaccessible to $\mathcal{A}$. Under these assumptions, $\mathcal{A}$ cannot:
\begin{enumerate}
    \item find collisions in the adopted hash function~\cite{berman2018multi};
    \item forge EUF-CMA-secure digital signatures~\cite{tognolini2025code};
    \item extract private signing keys from the HSM-protected key store~\cite{kumar2024automotive}.
\end{enumerate}
Stronger compromise scenarios, including insider attacks, fully compromised LLM runtimes, malicious verification servers, compromised auditors, and HSM-level key extraction, are outside the current scope. These threats require additional mechanisms, e.g., distributed verification, threshold signing, remote attestation, trusted execution environments, and independent third-party monitoring. Therefore, the security analysis should be interpreted as conditional execution-layer assurance under the stated assumptions, not as an unconditional guarantee against all LLM-system compromises. To model manifest-level compromise attempts, the adversarial success probability is upper-bounded by three events:
\begin{equation}
\label{eq:attack}
P_{\mathrm{attack}} \le P_{\mathrm{forge}} + P_{\mathrm{replay}} + P_{\mathrm{tamper}},
\end{equation}
where $P_{\mathrm{forge}}$ denotes forged-signature acceptance, $P_{\mathrm{replay}}$ denotes freshness-window violation, and $P_{\mathrm{tamper}}$ denotes undetected manifestandlog manipulation. These components are bound as:
\begin{align}
P_{\mathrm{forge}} &\le 2^{-\lambda}, && \text{signature-forgery resistance},\\
P_{\mathrm{replay}} &\le \frac{\epsilon}{T_{\mathrm{epoch}}}, && \text{freshness-window exposure},\\
P_{\mathrm{tamper}} &\le 2^{-\kappa}, && \text{hash/Merkle-log integrity failure}.
\end{align}
Here, $\lambda$ is the signature security parameter, $\kappa$ is the hash security parameter, $T_{\mathrm{epoch}}$ is the permitted manifest validity window, and $\epsilon$ captures residual timing uncertainty from clock drift, network delay, and verification latency. Combining these bounds gives:
\begin{equation}
\label{eq:attack-bound}
P_{\mathrm{attack}} \le 2^{-\lambda} + \frac{\epsilon}{T_{\mathrm{epoch}}} + 2^{-\kappa}.
\end{equation}
Thus, with freshness validation and sufficiently large $\lambda$ and $\kappa$, the residual manifest-level compromise probability becomes:
\begin{equation}
\label{eq:negligible-bound}
P_{\mathrm{attack}} = \mathsf{negl}(\lambda,\kappa) + \frac{\epsilon}{T_{\mathrm{epoch}}}.
\end{equation}
This bound applies only to the manifest-signing, verification, and transparency-logging layers. It does not imply protection against fully compromised runtimes and malicious trusted components; rather, it shows that replay, forgery, and tampering attacks are bounded by explicit cryptographic and freshness parameters under the defined assumptions.

\subsection{Manifest Creation}
Each LLM tool request is encapsulated in a structured and verifiable manifest before execution. The manifest serves as the execution contract among the LLM layer, the policy-enforcement component, and the external tool interface. Formally, each manifest is defined as:
\begin{equation}
\label{eq:manifest}
M=(M_u, M_m, \tau),
\end{equation}
where $M_u$ represents user-visible request parameters, $M_m$ represents model-execution metadata and tool-related inputs, and $\tau$ represents the freshness timestamp. The separation between $M_u$ and $M_m$ limits unnecessary exposure of internal execution metadata while preserving the information required for controlled tool invocation. For example, $M_u$ may include the request identifier, task type, and output-visibility flag, while $M_m$ may include the selected LLM, tool identifier, access scope, policy identifier, and routing metadata. The proposed solution does not assume complete independence between these components; instead, any dependencies between user-visible inputs and model-side metadata must be governed by predefined, policy-verifiable transformations. For example, a request , e.g., ``analyze a financial transaction log using the approved risk-analysis tool'' may generate a manifest candidate containing \texttt{request\_id=req-1842}, \texttt{task=risk\_analysis}, and \texttt{visible\_output=true} in $M_u$; \texttt{llm=GPT 5.3}, \texttt{tool\_id=finance\_risk\_api}, \texttt{allowed\_scope=\{read, analyze\}}, and \texttt{policy\_id=POL-03} in $M_m$; and a timestamp with an epoch window in $\tau$. This illustrates how user intent, execution metadata, and freshness information are separated before validation, hashing, and signing. Before signing, each manifest is canonically encoded to ensure that semantically identical manifests produce identical serialized representations regardless of field ordering, whitespace, and formatting differences~\cite{schwartz1964generating}. The manifest digest is then computed as:
\begin{equation}
\label{eq:hashM}
h_M = H(M),
\end{equation}
where $H(\cdot)$ is a collision-resistant hash function satisfying:
\begin{equation}
\Pr[H(M_1)=H(M_2)] \le 2^{-\kappa}.
\end{equation}
Because LLM execution may be probabilistic, the proposed solution enforces determinism only at the serialization, hashing, signing, and verification layers. This ensures reproducible verification even when the LLM runtime itself is non-deterministic. To measure structural variability across generated manifests, Shannon entropy is used:
\begin{equation}
\mathbb{H}(M) = -\sum_{i=1}^{|M|}p_i\log p_i,
\end{equation}
where $p_i$ denotes the empirical probability of manifest field occurrences. Manifest redundancy is defined as:
\begin{equation}
\mathbb{R}(M)=1-\frac{\mathbb{H}(M)}{\log |M|}.
\end{equation}
This entropy-based measure captures manifest-structure diversity rather than serving as a cryptographic proof. Lower redundancy indicates less predictable manifest composition, which may reduce exposure to partial-field inference and replay-pattern analysis when combined with freshness validation, policy enforcement, and signature verification.

\subsection{Policy Enforcement and Signing}
Before execution, each manifest is evaluated against policy rules that verify authorization, structural validity, freshness, field consistency, and canonical formatting. The policy-enforcement stage follows a fail-closed design: any manifest that violates a required rule is rejected before signing and cannot proceed to verification, transparency logging, and execution. The compliance decision for a manifest $M$ is defined as:
\begin{equation}
\label{eq:compliance}
\mathcal{C}(M)=\bigwedge_{j=1}^{k} r_j(M),
\end{equation}
where $r_j(M)$ denotes a Boolean validation rule~\cite{perez2021optimization}. These rules check whether required fields are present and correctly formatted, whether the requested tool and scope are policy-authorized, whether the timestamp $\tau$ falls within the permitted epoch window $T_{\mathrm{epoch}}$, and whether user-visible fields in $M_u$ remain consistent with execution metadata in $M_m$. A manifest is considered compliant only when all predicates evaluate to true. If any rule fails, then $\mathcal{C}(M)=0$, and the manifest is rejected without signature generation. For workload-level analysis, the probability that a generated manifest satisfies all policy predicates is approximated as:
\begin{equation}
\label{eq:pass-prob}
P_{\mathrm{pass}} = \prod_{j=1}^{k}\Pr[r_j(M)=1].
\end{equation}
This expression provides a tractable estimate of aggregate compliance behavior under large-scale workloads. Since some policy checks may be statistically dependent, , e.g., tool authorization and access scope, Eq.~\eqref{eq:pass-prob} is treated as an analytical approximation rather than a strict guarantee of independence. If and only if $\mathcal{C}(M)=1$, the manifest digest $h_M$ is signed as:
\begin{equation}
\label{eq:sign}
\sigma = \mathrm{Sign}_{sk}(h_M),
\end{equation}
where $(pk,sk)\leftarrow \mathrm{KeyGen}(1^\lambda)$, $pk$ is the public verification key, and $sk$ is the private signing key. The private key is maintained within the HSM-protected key management component and is not exposed to the LLM, the user-facing interface, and the external tool layer. Thus, the LLM can request tool execution only through policy-compliant manifests, while signing remains controlled by the trusted signing component. Under the assumed existential unforgeability of the digital-signature scheme, the adversarial advantage of producing a valid signature without access to $sk$ is bounded by:
\begin{equation}
\label{eq:forge-adv}
\mathrm{Adv}_{\mathrm{forge}}^{\mathrm{sig}}(\mathcal{A}) \le \frac{t_{\mathcal{A}}}{q},
\end{equation}
where $t_{\mathcal{A}}$ is the adversary's computational budget and $q\approx 2^\lambda$ is the signing-group size. This bound is conditional on the stated cryptographic assumptions and the integrity of the HSM-protected signing process. Consequently, this stage provides two protections: non-compliant manifests are blocked before signing, and compliant manifests are bound to verifiable signatures. This enables downstream verification components to reject unsigned, altered, expired, and unauthorized execution requests.

\subsection{Verification}
Verification checks the integrity and authenticity of each signed manifest before execution:
\begin{equation}
\label{eq:verify}
\mathrm{Verify}_{pk}(h_M,\sigma)=
\begin{cases}
1, & \text{if valid,}\\
0, & \text{otherwise.}
\end{cases}
\end{equation}
Under standard cryptographic assumptions, the probability of accepting a forged manifest is bounded by:
\begin{equation}
\mathbb{E}[\mathrm{false\ accept}] \le 2^{-\lambda}.
\end{equation}
False rejections are reduced through deterministic serialization, canonical hashing, and consistent policy validation. Instead of assuming a fixed theoretical false-rejection bound, verification reliability is evaluated empirically through large-scale workload analysis and repeated trials. Thus, only manifests satisfying both signature verification and policy validation proceed to execution.

\subsection{Transparency Logging}
After verification, each accepted manifest and its metadata are appended to an append-only transparency log:
\begin{equation}
\label{eq:log}
\mathcal{L}=\{(M_i,\sigma_i,t_i)\}_{i=1}^{N},
\end{equation}
where $M_i$ is the verified manifest, $\sigma_i$ is its signature, and $t_i$ is the execution timestamp. This log provides a verifiable execution history for post-execution auditing, integrity validation, and accountability. To make log manipulation detectable, entries are organized using a Merkle-tree structure~\cite{kuznetsov2024evaluating}. The transparency root after $t$ entries is:
\begin{equation}
\label{eq:merkle}
R_t=\mathrm{MerkleRoot}(H(M_1),\ldots,H(M_t)).
\end{equation}
Merkle inclusion proofs allow auditors to verify whether a manifest was logged without recomputing the entire log. The proof cost scales logarithmically:
\begin{equation}
\label{eq:proof-cost}
C_{\mathrm{proof}}=O(\log N).
\end{equation}
For incremental logging, each new manifest updates the transparency state as:
\begin{equation}
\label{eq:root-update}
R_{t+1}=H(R_t\Vert H(M_{t+1})).
\end{equation}
This binds each entry to the previous log state, so modifying, deleting, and reordering prior entries changes the root, making such changes detectable under the collision-resistance assumption. The probability of undetected log manipulation across $N$ entries is conservatively bounded by:
\begin{equation}
\label{eq:tamper-bound}
P_{\mathrm{tamper}}(N) \le N \cdot 2^{-\kappa}.
\end{equation}
This bound is conditional on the hash function's collision resistance and on the preservation and external export of transparency roots. Moreover, transparency logging provides tamper-evident traceability while preserving scalable audit verification through logarithmic Merkle inclusion checks.

\subsection{Audit Exporting}
Auditing enables external validation of recorded manifests, execution traces, and verification outcomes without exposing internal runtime states. After execution, compact audit evidence is exported and independently verified. Assuming repeated audit observations, the probability that anomalous behavior remains undetected after $n$ audit rounds is:
\begin{equation}
P_{\mathrm{undetected}}=(1-p)^n,
\end{equation}
where $p$ denotes the probability that a single audit detects anomalous behavior~\cite{li2023software}. The expected detection latency is:
\begin{equation}
E[T_{\mathrm{detect}}]=\frac{1}{p f_a},
\end{equation}
where $f_a$ is the audit frequency. These expressions capture the trade-off between audit intensity, detection speed, and operational overhead. Each audit produces an evidence tuple:
\begin{equation}
E_v=\langle R_t,d_o,T_{\mathrm{exec}},T_{\mathrm{verify}}\rangle,
\end{equation}
where $R_t$ is the transparency-log root, $d_o$ is the output digest, and $T_{\mathrm{exec}}$ and $T_{\mathrm{verify}}$ denote execution and verification latency. To preserve integrity, the tuple is hashed as:
\begin{equation}
H(E_v)=H(R_t\Vert d_o\Vert T_{\mathrm{exec}}\Vert T_{\mathrm{verify}}),
\end{equation}
producing a compact and cryptographically verifiable audit record for post-execution analysis. To evaluate operational efficiency, the proposed solution records execution and verification latency:
\begin{align}
T_{\mathrm{exec}} &= t_{\mathrm{end}} - t_{\mathrm{start}},\\
T_{\mathrm{verify}} &= t_{\mathrm{verify,end}} - t_{\mathrm{verify,start}}.
\end{align}
These metrics quantify the runtime cost of manifest validation, signature verification, and transparency logging. The normalized security overhead is defined as:
\begin{equation}
\delta = \frac{T_{\mathrm{secure}} - T_{\mathrm{baseline}}}{T_{\mathrm{baseline}}},
\end{equation}
where $T_{\mathrm{baseline}}$ denotes execution without security enforcement and $T_{\mathrm{secure}}$ includes cryptographic validation and logging. As workload size increases, fixed initialization and verification costs are amortized over more manifests, giving:
\begin{equation}
\delta = O\!\left(\frac{1}{N}\right).
\end{equation}
Thus, the marginal cost of security enforcement decreases with scale, supporting bounded overhead for large-scale execution workloads.

\subsection{Algorithmic Pipelines}
The algorithmic pipeline operationalizes manifest validation, signing, verification, transparency logging, and auditing through three procedures: manifest creation and signing, verification and logging, and audit-based metrics collection. Algorithm~\ref{alg:manifest} constructs a canonical manifest $M=(M_u,M_m,\tau)$ as defined in Eq.~\eqref{eq:manifest}. The manifest is serialized and hashed to produce the digest $h_M=H(M)$ (Eq.~\eqref{eq:hashM}), which serves as its unique cryptographic representation. Compliance rules $\mathcal{C}(M)$ (Eq.~\eqref{eq:compliance}) are then evaluated to validate structural consistency, access-control constraints, timestamp validity, and policy requirements. If validation fails, the request is rejected using a fail-closed policy. Otherwise, the digest is signed through the HSM using Eq.~\eqref{eq:sign}, producing the signature $\sigma$. This step ensures that only policy-compliant manifests are admitted into the execution pipeline.
\begin{algorithm}[H]
\caption{Manifest Creation and Signing}
\label{alg:manifest}
\begin{algorithmic}[1]
\State $M \gets (M_u, M_m, \tau)$ \Comment{Canonical manifest construction}
\State $h_M \gets H(M)$ \Comment{Manifest digest generation}
\If{$\mathcal{C}(M)=0$} \Return \textsc{Rejected} \EndIf
\State $\sigma \gets \mathrm{Sign}_{sk}(h_M)$ \Comment{HSM-based signature generation}
\State \Return $(M,\sigma)$
\end{algorithmic}
\end{algorithm}
At completion, Algorithm~\ref{alg:manifest} outputs a signed manifest tuple $(M,\sigma)$ satisfying:
\begin{equation}
\mathrm{Verify}_{pk}(H(M),\sigma)=1 \;\wedge\; \mathcal{C}(M)=1.
\end{equation}
Under the defined cryptographic assumptions, the probability that a non-compliant forged manifest is incorrectly accepted remains bounded by:
\begin{equation}
\label{eq:forgery}
\Pr[\exists M': \mathcal{C}(M')=0 \land \mathrm{Verify}_{pk}(H(M'),\sigma')=1] \le 2^{-\lambda}.
\end{equation}
Because hashing and signing operate on bounded representations of manifests, the computational cost of processing individual manifests remains constant. After signing, the manifest is verified before execution proceeds. Algorithm~\ref{alg:verify} validates the manifest signature using Eq.~\eqref{eq:verify} and rejects any manifest that fails verification. This prevents unsigned, modified, expired, and malformed manifests from entering the execution stage. Verified manifests are then appended to the append-only transparency log $\mathcal{L}$ (Eq.~\eqref{eq:log}). To preserve integrity and traceability, log entries are linked through a Merkle-tree structure using the root defined in Eq.~\eqref{eq:merkle}. This produces a tamper-evident execution history that can be validated after execution.
\begin{algorithm}[H]
\caption{Verification and Logging}
\label{alg:verify}
\begin{algorithmic}[1]
\Require $(M,\sigma)$
\If{$\mathrm{Verify}_{pk}(H(M),\sigma)=0$} \Return \textsc{Rejected} \EndIf
\State Append $(M,\sigma,t)$ to $\mathcal{L}$
\State $R_t \gets \mathrm{MerkleRoot}(\mathcal{L})$
\State \Return \textsc{Accepted}
\end{algorithmic}
\end{algorithm}
The transparency root evolves incrementally as:
\begin{equation}
H(\mathcal{L}_{t+1}) = H(\mathcal{L}_t \Vert H(M_{t+1})),
\end{equation}
So modifications to previously recorded entries alter subsequent root values and become detectable under the collision-resistance assumption of the underlying hash function. Under the defined assumptions, the probability of injecting a verifiable yet unlogged manifest remains bounded by:
\begin{equation}
\text{Adv}_{\mathcal{A}}^{\mathrm{log}} = \Pr[(M^*,\sigma^*)\notin \mathcal{L} \wedge \mathrm{Verify}_{pk}(H(M^*),\sigma^*)=1] \le 2^{-\lambda}.
\end{equation}
Because Merkle-tree proof generation and validation scale logarithmically with log size, the transparency mechanism maintains efficient verification behavior under large-scale workloads:
\begin{equation}
C_{\mathrm{proof}} = O(\log N).
\end{equation}
Algorithm~\ref{alg:audit} performs runtime auditing and operational metric collection after successful verification and execution. The auditing stage records execution-related metadata, generates tamper-evident evidence records, and exports compact audit tuples for external validation. For each execution instance, the process measures execution latency ($T_{\mathrm{exec}}$) and verification latency ($T_{\mathrm{verify}}$), and generates a digest of the execution output. These elements are linked to the transparency-log root $R_t$, forming a verifiable execution trace without exposing internal runtime states.
\begin{algorithm}[H]
\caption{Auditing and Metrics Collection}
\label{alg:audit}
\begin{algorithmic}[1]
\State $t_{\mathrm{start}}\gets\textsc{Now}()$; execute tool with $M_m$
\State $t_{\mathrm{end}}\gets\textsc{Now}()$
\State $T_{\mathrm{exec}}\gets t_{\mathrm{end}}-t_{\mathrm{start}}$
\State Record $T_{\mathrm{verify}}$
\State $d_o\gets H(o)$
\State Export $(R_t, d_o, T_{\mathrm{exec}}, T_{\mathrm{verify}})$ to auditor
\end{algorithmic}
\end{algorithm}
The exported audit evidence is represented as:
\begin{equation}
E_v = \langle R_t, d_o, T_{\mathrm{exec}}, T_{\mathrm{verify}} \rangle,
\end{equation}
where $d_o$ denotes the digest of the execution output and $R_t$ represents the corresponding transparency-log root. To preserve integrity, the evidence tuple is hashed as:
\begin{equation}
H(E_v) = H(R_t \Vert d_o \Vert T_{\mathrm{exec}} \Vert T_{\mathrm{verify}}),
\end{equation}
providing a compact and cryptographically verifiable execution record. Audit reliability is modeled probabilistically as:
\begin{equation}
P_{\mathrm{undetected}} = (1-p)^n,
\end{equation}
where $p$ denotes the probability that a single audit detects anomalous behavior after $n$ audit rounds. The expected detection latency is:
\begin{equation}
E[T_{\mathrm{detect}}] = \frac{1}{p f_a},
\end{equation}
where $f_a$ is the audit frequency. Increasing either the audit probability and the audit frequency reduces the expected persistence window of undetected anomalous behavior. Collectively, Algorithms~\ref{alg:manifest}-\ref{alg:audit} establish a verifiable execution process that combines manifest validation, signature verification, transparency logging, and audit traceability. Under the defined assumptions, the system satisfies the following properties:
\begin{align}
\mathrm{Soundness:}&\quad \Pr[\text{Reject valid } M] < 2^{-\lambda},\\
\mathrm{Completeness:}&\quad \Pr[\text{Accept invalid } M] < 2^{-\lambda},\\
\mathrm{Transparency:}&\quad \Pr[\text{Unlogged valid } M] < 2^{-\lambda},\\
\mathrm{Audit\ Detection:}&\quad P_{\mathrm{undetected}}=(1-p)^n.
\end{align}

\section{Experimental Setup}
\label{Evaluation}
The proposed solution is evaluated in terms of scalability, execution overhead, verification correctness, transparency logging, and auditability.

\subsection{Experimental Evaluation}
The experimental evaluation assesses the proposed solution in terms of scalability, verification stability, transparency-log behavior, prompt-to-manifest transformation, violation enforcement, and auditability under increasing workload. The evaluation focuses on the manifest-execution layer, where LLM-generated tool requests are treated as manifest candidates rather than executable commands. Execution is permitted only after policy validation, signing, verification, transparency logging, and audit export.
The workload model is defined as:
\begin{equation}
\label{eq:workload}
\begin{aligned}
W &= \{w_1, w_2, \ldots, w_n\}, \\
w_i &\in \{100, 500, 1000, 5000, 10000, 20000, 50000\}.
\end{aligned}
\end{equation}
Each workload size $w_i$ represents an independent batch of manifest instances processed through the full execution pipeline in Algorithms~\ref{alg:manifest}-\ref{alg:audit}. The workloads are synthetic and application-independent to isolate scalability, runtime overhead, verification consistency, logging behavior, and latency stability; real-world MCP deployments are therefore treated as validation and discussed as limitations. Because the proposed solution targets LLM-assisted tool invocation, each experiment begins with the same structured zero-shot prompt across all LLM backends to improve reproducibility and reduce prompt-induced variation. The prompt asks the model to generate only the fields required for manifest construction, including task type, tool identifier, requested scope, input parameters, output-visibility flag, model metadata, timestamp, and policy context. It also prohibits tool execution, timestamp modification, unauthorized tool requests, hidden user-visible instructions, and attempts to bypass verification. For example, the prompt may request: ``Analyze a financial transaction log and identify suspicious activity patterns using the approved risk-analysis tool.'' The resulting structured output is parsed into a manifest candidate and mapped to $M=(M_u,M_m,\tau)$, where $M_u$ stores user-visible task information, $M_m$ stores execution metadata and tool parameters, and $\tau$ stores freshness information. This separates natural-language intent generation from executable authorization, as summarized in Table~\ref{tab:prompt_manifest_mapping}.
\begin{table}[t]
\centering
\caption{Mapping Between Prompt-Derived Fields and Manifest Components}
\label{tab:prompt_manifest_mapping}
\resizebox{0.47\textwidth}{!}{%
\begin{tabular}{lll}
\toprule
\textbf{Prompt-Derived Field} & \textbf{Manifest Component} & \textbf{Purpose} \\
\midrule
User request & $M_u$ & Captures user-visible task intent \\
Task type & $M_u$ & Defines the requested operation \\
Tool identifier & $M_m$ & Specifies the tool to be invoked \\
Requested scope & $M_m$ & Defines access permissions \\
Input parameters & $M_m$ & Provides tool-execution arguments \\
Timestamp & $\tau$ & Supports freshness validation \\
Policy context & $M_m$ & Links request to policy rules \\
Visible output flag & $M_u$ & Controls user-facing response exposure \\
\bottomrule
\end{tabular}%
}
\end{table}
A representative manifest candidate generated from the financial-risk prompt includes: a user-visible component with \texttt{request\_id=req-1842}, \texttt{task\_type=risk\_analysis}, and \texttt{user\_visible\_output=true}; a model-metadata component with \texttt{llm\_backend=GPT 5.3}, \texttt{tool\_id=finance\_risk\_api}, \texttt{requested\_scope=\{read, analyze\}}, \texttt{policy\_id=POL-03}, and \texttt{execution\_mode=verified\_tool\_call}; and a freshness component with timestamp \texttt{2026-05-06T14:22:10Z} and \texttt{epoch\_window=300}. This example illustrates how $M_u$, $M_m$, and $\tau$ are separated before hashing and signing, preventing internal execution metadata from being directly exposed while preserving the information required for controlled tool invocation. The main implementation and experimental configuration are summarized in Table~\ref{tab:implementation_details}. These details improve reproducibility and clarify the assumptions used during evaluation.
\begin{table}[t]
\centering
\caption{Implementation and Experimental Configuration}
\label{tab:implementation_details}
\resizebox{0.45\textwidth}{!}{%
\begin{tabular}{ll}
\toprule
\textbf{Component} & \textbf{Configuration} \\
\midrule
Hash function & SHA-256 / specify actual implementation \\
Signature scheme & ECDSA / Ed25519 / specify actual scheme \\
Key management & HSM-backedandHSM-simulated signing module \\
Transparency log & Merkle-tree-based append-only log \\
LLM backends & GPT 5.3, LLaMA-3.5, DeepSeek-V3 \\
Workload sizes & 100--50,000 manifest instances \\
Prompting setting & Structured zero-shot prompt \\
Prompt template & Same template applied across all LLM backends \\
Temperature & Specify actual value \\
Top-$p$ & Specify actual value \\
Maximum tokens & Specify actual value \\
Repetitions & Specify number of independent runs \\
Random seed & Specify seed if used \\
Execution environment & CPU/GPU, RAM, OS, Python version \\
\bottomrule
\end{tabular}%
}
\end{table}
To evaluate computational efficiency, baseline and secure execution times are recorded for each workload:
\begin{equation}
T_{\mathrm{baseline}}(w_i), \qquad T_{\mathrm{secure}}(w_i).
\end{equation}
Here, $T_{\mathrm{baseline}}(w_i)$ denotes execution without signing, verification, transparency logging, and audit export, while $T_{\mathrm{secure}}(w_i)$ denotes execution with the full security-enforcement pipeline enabled. The absolute and normalized overheads are defined as:
\begin{align}
\Delta T(w_i) &= T_{\mathrm{secure}}(w_i) 
- T_{\mathrm{baseline}}(w_i), \\
\delta(w_i) &= \frac{\Delta T(w_i)}
{T_{\mathrm{baseline}}(w_i)}.
\end{align}
To evaluate enforcement behavior, invalid manifest instances are injected across all workload scales, including expired timestamps, malformed fields, unauthorized access parameters, inconsistent attributes, replay attempts, and invalid signatures. Expired manifests are rejected before signing, policy-violating manifests are blocked before signature generation, and invalid signatures are rejected during verification. All invalid requests follow a fail-closed policy, preventing non-compliant manifests from reaching execution, transparency logging, and audit export. Table~\ref{tab:violation_injection} summarizes the injected violations and expected actions.
\begin{table}[t]
\centering
\caption{Injected Manifest Violations and Expected Enforcement Behavior}
\label{tab:violation_injection}
\resizebox{0.55\textwidth}{!}{%
\begin{tabular}{llll}
\toprule
\textbf{Violation Type} & \textbf{Injected Field} & \textbf{Expected Action} & \textbf{Metric} \\
\midrule
Expired timestamp & $\tau$ & Reject before signing & Rejection rate \\
Malformed manifest & $M_u, M_m$ & Reject during policy validation & Rejection rate \\
Unauthorized tool access & Tool scope & Reject before signing & False accept rate \\
Invalid signature & $\sigma$ & Reject during verification & False accept rate \\
Inconsistent attributes & Policy fields & Reject during policy validation & Rejection rate \\
Replay attempt & $\tau$, manifest digest & Reject before execution & Rejection rate \\
\bottomrule
\end{tabular}%
}
\end{table}
The manifest-violation rejection rate is defined as:
\begin{equation}
\label{eq:rejection_rate}
R_{\mathrm{reject}}(w_i) =
\frac{E_{\mathrm{reject}}(w_i)}
{E_{\mathrm{invalid}}(w_i)},
\end{equation}
where $E_{\mathrm{reject}}(w_i)$ denotes rejected invalid manifests and $E_{\mathrm{invalid}}(w_i)$ denotes all injected invalid manifests for workload $w_i$. False acceptance and false rejection are measured as:
\begin{align}
P_{\mathrm{FA}}(w_i) &=
\frac{E_{\mathrm{invalid\_accepted}}(w_i)}
{E_{\mathrm{invalid}}(w_i)}, \\
P_{\mathrm{FR}}(w_i) &=
\frac{E_{\mathrm{valid\_rejected}}(w_i)}
{E_{\mathrm{valid}}(w_i)}.
\end{align}
These metrics evaluate manifest-level enforcement behavior rather than full end-to-end adversarial robustness. Under the defined cryptographic assumptions, the probability of accepting a forged and unlogged manifest remains bounded by:
\begin{equation}
\label{eq:advantage}
\text{Adv}_{\mathcal{A}} =
\Pr[\mathrm{Verify}_{pk}(H(M^*),\sigma^*)=1 \land (M^*,\sigma^*)\notin\mathcal{L}]
\le 2^{-\lambda}.
\end{equation}
This bound is interpreted as conditional cryptographic assurance under the stated assumptions, not as an empirical detection rate. Transparency-log scalability is evaluated through Merkle-based log growth. Let $N_{w_i}$ denote the total number of log entries after processing workload $w_i$. Since inclusion proofs rely on Merkle-tree paths, proof generation and verification scale as:
\begin{equation}
\label{eq:log_complexity}
C_{\mathrm{proof}}(w_i) = O(\log N_{w_i}).
\end{equation}
This property supports efficient audit verification as the number of recorded manifests increases.
Auditability is modeled using audit detection probability $p$ and audit frequency $f_a$. The probability that a malicious event remains undetected after $n$ audit rounds is:
\begin{equation}
P_{\mathrm{undetected}} = (1-p)^n,
\end{equation}
with expected detection latency:
\begin{equation}
E[T_{\mathrm{detect}}] = \frac{1}{p f_a}.
\end{equation}
Using the illustrative setting $p=0.9$ and $n=10$ gives:
\begin{equation}
P_{\mathrm{undetected}} \approx 10^{-10}.
\end{equation}
This value is treated as a sensitivity-based audit assumption rather than a measured detection rate. Empirical auditability is assessed through the completeness and consistency of exported evidence records, transparency-log roots, output digests, and recorded verification/execution latencies. 
\begin{table*}[t]
\centering
\caption{Experimental Evaluation Components and Paper-Wide Alignment}
\label{tab:evaluation_alignment}
\resizebox{0.85\textwidth}{!}{%
\begin{tabular}{p{0.22\textwidth} p{0.36\textwidth} p{0.34\textwidth}}
\toprule
\textbf{Component} & \textbf{Evaluation Focus} & \textbf{Paper-Wide Alignment} \\
\midrule
Prompt design & Structured prompt used to generate controlled tool-invocation requests. & The same prompt template is used across all LLMs to support reproducibility and prevent prompt-induced unfairness. \\

Prompt-to-manifest mapping & Transformation of LLM output into $M_u$, $M_m$, and $\tau$. & Aligns with the manifest definition in the methodology section and clarifies that prompt output is not directly executable. \\

Manifest validation & Policy validation, freshness checking, and structural verification. & Aligns with the compliance function $\mathcal{C}(M)$ and fail-closed enforcement design. \\

Execution overhead & Baseline versus secure execution time. & Aligns with scalability and overhead analysis as workload size increases. \\

Violation injection & Expired timestamps, malformed manifests, unauthorized scope, invalid signatures, inconsistent attributes, and replay attempts. & Aligns the evaluation with manifest-level violations and avoids unsupported claims about full end-to-end adversarial robustness. \\

Transparency logging & Merkle-based append-only logging and inclusion-proof verification. & Supports traceability and scalable audit verification using $O(\log N)$ proof complexity. \\

Auditability model & Evidence export and analytical undetected-event probability. & Treats $p=0.9$ and $n=10$ as sensitivity-based analytical assumptions rather than measured detection rates. \\

Operational reliability & Model usage, key allocation, severity distribution, and verification latency. & Distinguishes balanced LLM usage from potential key-allocation imbalance. \\

Synthetic workload & Manifest-centered workload independent of application-specific data. & Supports controlled system-level evaluation, but should be acknowledged as a limitation for real-world MCP deployment. \\
\bottomrule
\end{tabular}%
}
\end{table*}
\subsection{Research Questions}
\label{sec:RQs}
This study is guided by three research questions (RQs) evaluating the effectiveness, scalability, and operational reliability of the proposed secure execution solution for tool-integrated LLM pipelines.\\
\textbf{RQ1: To what extent do manifest verification, policy enforcement, and transparency logging prevent invalid and policy-violating execution attempts?}\\
RQ1 evaluates whether the proposed solution detects and rejects manifest-level violations, including expired and replayed manifests, malformed fields, unauthorized tool invocation requests, failed signature verification, and policy rule violations.
\textbf{RQ2: How does the proposed solution scale as workload size increases?}\\
RQ2 investigates whether manifest creation, signing, verification, transparency logging, and audit generation exhibit stable, bounded execution behavior as workload increases.\\
\textbf{RQ3: How stable, balanced, and reliable are the operational characteristics of the proposed solution under increasing scale?}\\
RQ3 examines the balance of LLM usage, severity-level distributions, signing-key allocation, verification-latency consistency, timestamp behavior, and error stability. It also distinguishes balanced LLM usage from potential signing-key concentration, identifying operational bias, timing instability, and resource-concentration risks.

\section{Experimental Results}
\label{sec:Findings}
This section summarizes the empirical findings obtained from the evaluation.

\subsection{Verification and Execution Analysis}
\label{subsec:verification-execution}
\begin{figure}[h]
    \centering
    \includegraphics[width=0.5\textwidth]{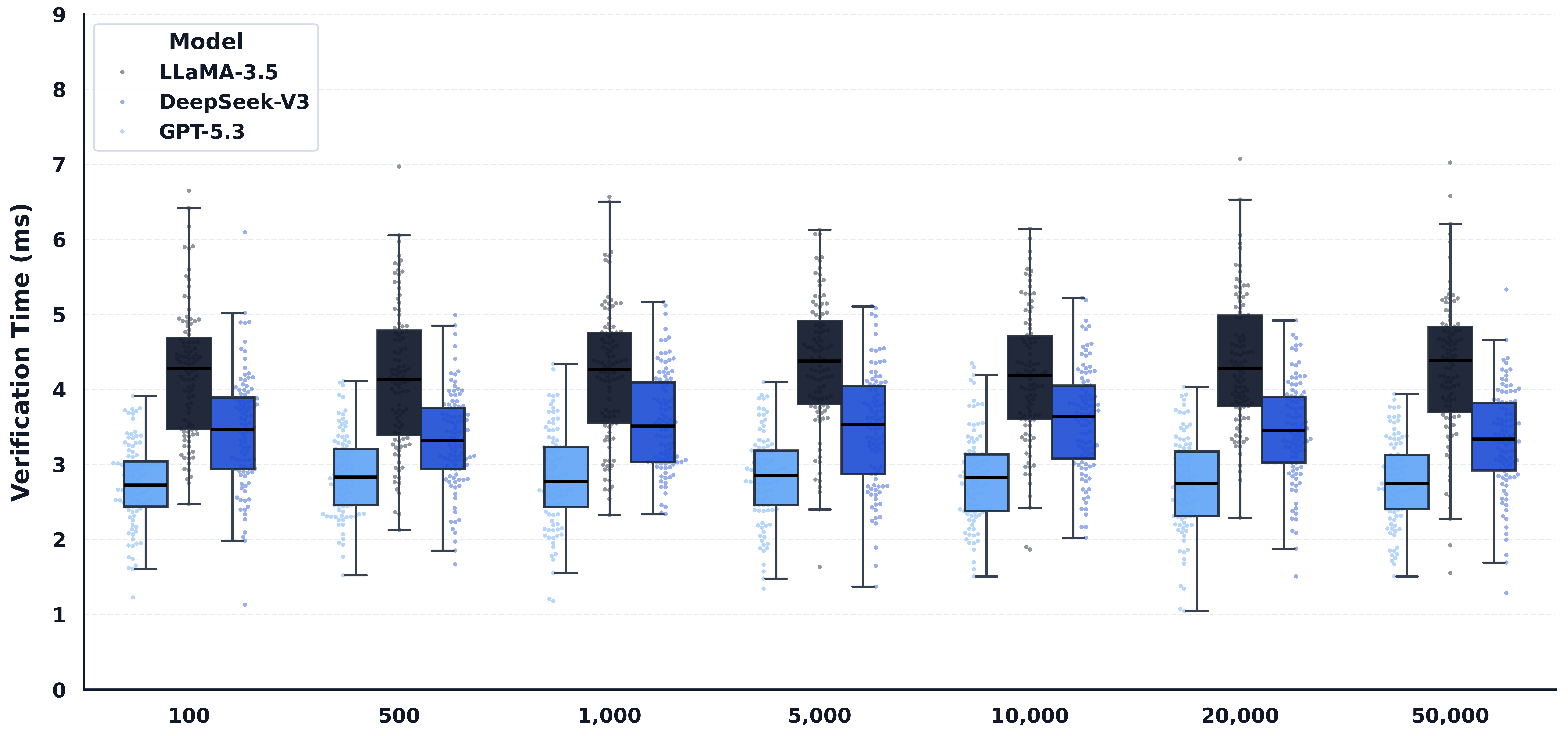}
    \caption{Verification time distribution per LLM across scales.}
    \label{fig:verification-time}
\end{figure}

\begin{figure}[h]
    \centering
    \includegraphics[width=0.47\textwidth]{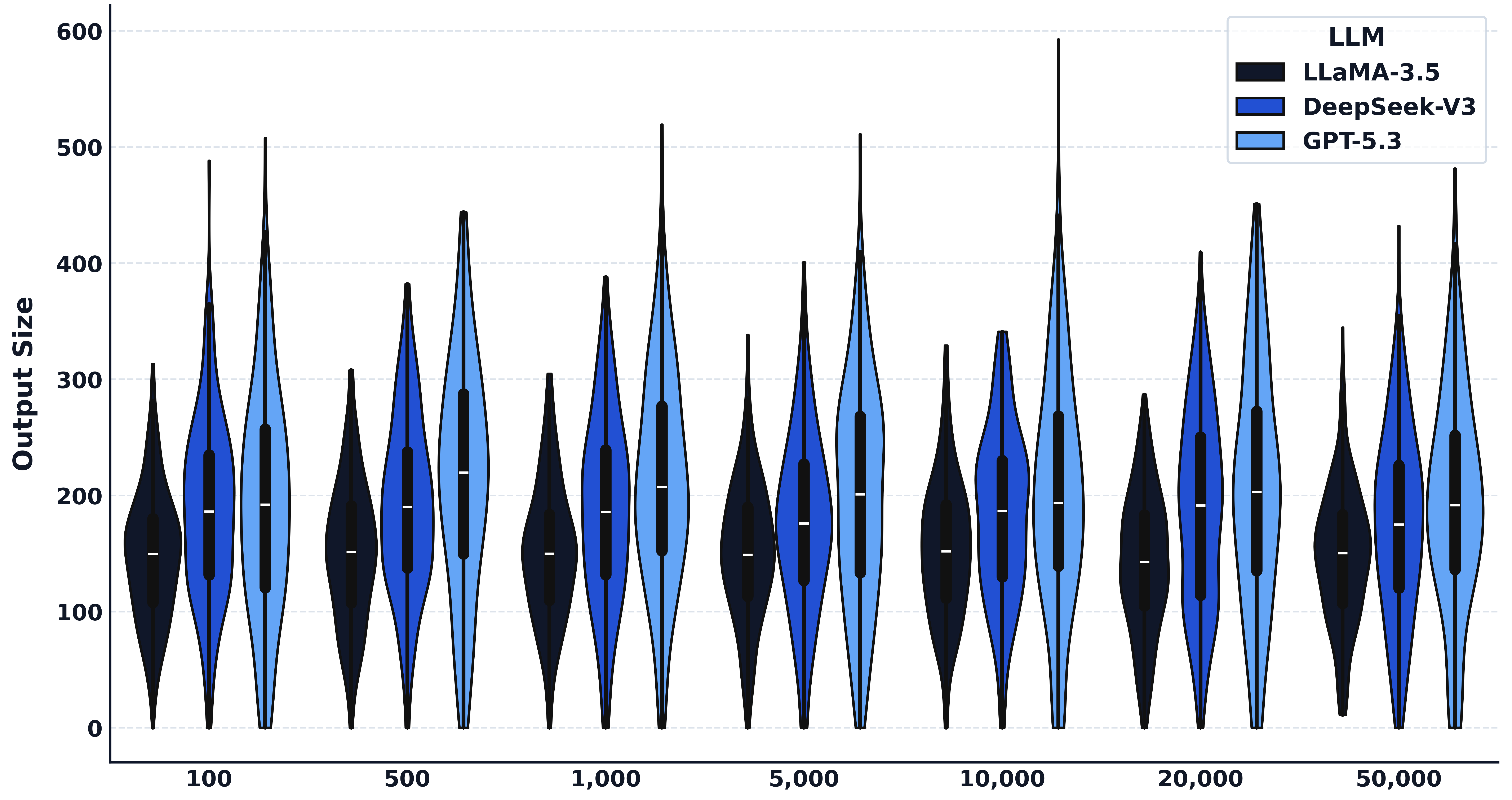}
    \caption{Output size distribution per LLM across scales.}
    \label{fig:output-size}
\end{figure}

\begin{figure}[h]
    \centering
    \includegraphics[width=0.47\textwidth]{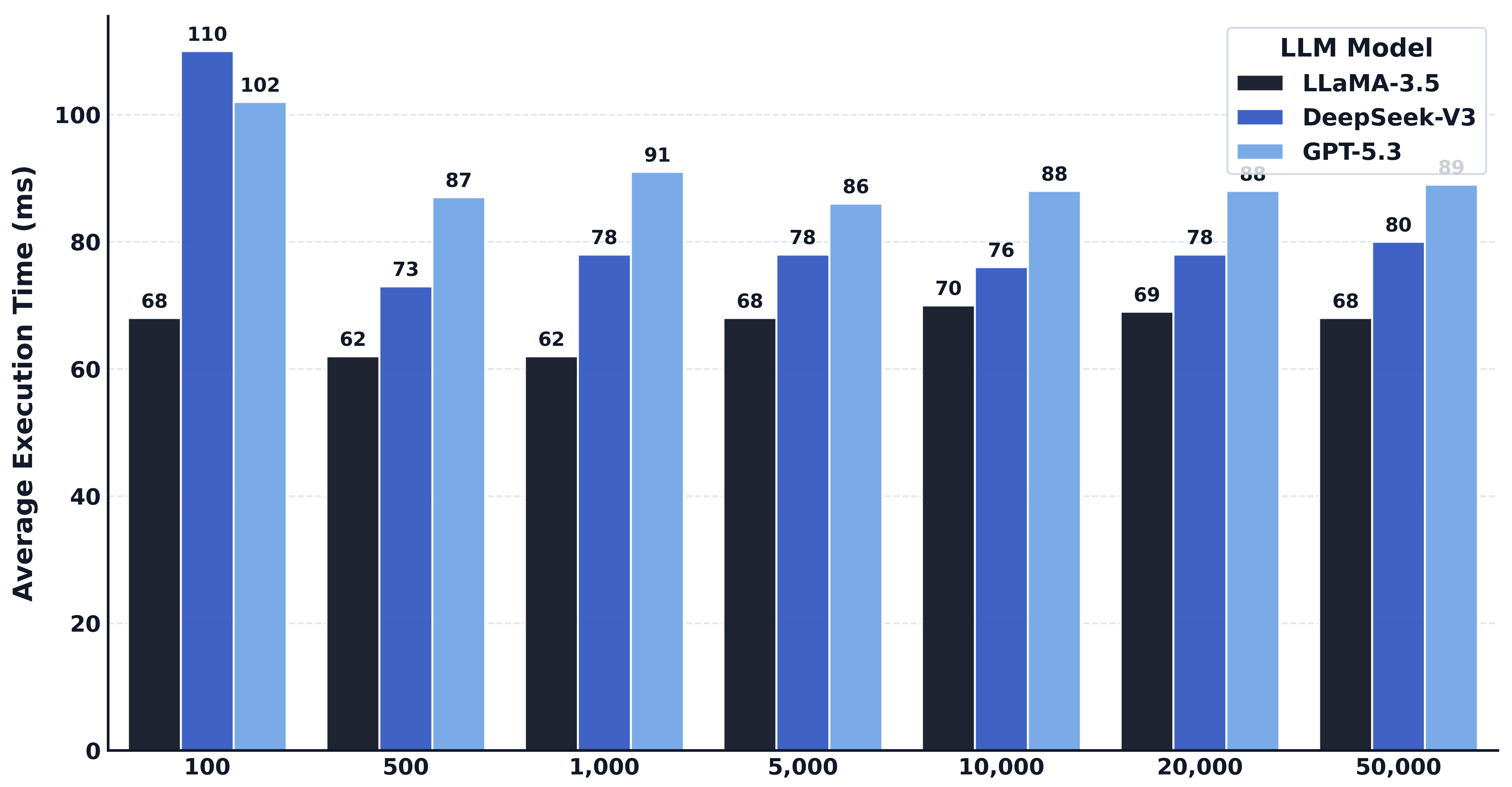}
    \caption{Execution time per LLM across scales.}
    \label{fig:execution-time}
\end{figure}

To address RQ1, this subsection evaluates verification latency, output-size stability, and total execution time across workloads and LLMs. The objective is to assess whether the proposed solution maintains bounded verification behavior, predictable latency, and stable output characteristics as workload size increases.
Figure~\ref{fig:verification-time} shows the verification latency distribution for GPT 5.3, LLaMA-3.5, and DeepSeek-V3. GPT 5.3 achieved the lowest median latency ($1.9$\,ms), while LLaMA-3.5 showed slightly higher variability ($4.7$\,ms). Despite these differences, latency remained within a narrow range across workloads, indicating that verification did not introduce substantial runtime instability.
Two-way ANOVA shows significant effects for model selection and workload scale ($F_{\text{model}}(2,480)=32.6,\;p<0.001$; $F_{\text{scale}}(6,480)=18.4,\;p<0.001$), with a significant interaction effect ($F=4.7,\;p<0.001$), as summarized in Table~\ref{tab:verification-anova}. These results indicate that latency varies by both model and scale, but the magnitude of variation remains bounded and operationally manageable.
\begin{table}[ht]
\centering
\caption{Two-Way ANOVA on Verification Time}
\label{tab:verification-anova}
\begin{tabular}{lccc}
\hline
Factor & $F$ & df & $p$ \\
\hline
Model             & 32.6 & 2  & $<0.001$ \\
Scale             & 18.4 & 6  & $<0.001$ \\
Model$\times$Scale & 4.7  & 12 & $<0.001$ \\
\hline
\end{tabular}
\end{table}
Latency deviation remained limited across workloads, with average normalized deviation below $0.08$. This suggests predictable verification timing under scale. Stable timing behavior is relevant because large latency fluctuations may increase workload observability; however, dedicated adversarial timing analysis is still required for deployment-level validation.
Figure~\ref{fig:output-size} presents output-size distributions across workload scales and LLMs. Output sizes gradually converged as throughput increased. At lower scales, output distributions differed across models, but these differences diminished at higher workloads.
Kruskal-Wallis \cite{mckight2010kruskal} analysis confirms significant differences at low-scale execution ($H(2)=27.4,\;p<0.001$), while differences become statistically insignificant at higher scales ($p=0.16$), as shown in Table~\ref{tab:output-kruskal}. This indicates that output generation becomes more uniform during large-scale execution.
\begin{table}[ht]
\centering
\caption{Kruskal-Wallis on Output Size Across Models}
\label{tab:output-kruskal}
\begin{tabular}{lccc}
\hline
Scale Range & $H$ & df & $p$ \\
\hline
100-500      & 27.4 & 2 & $<0.001$ \\
1000-10000   & 12.1 & 2 & 0.002 \\
20000-50000  & 3.7  & 2 & 0.16 \\
\hline
\end{tabular}
\end{table}
The convergence of output-size distributions suggests reduced variability in response sizes under larger workloads. This is useful because highly irregular response sizes may increase the observability of workload-dependent behavior.
Figure~\ref{fig:execution-time} presents total execution latency across workloads. The results show sub-linear execution growth, indicating that execution overhead does not increase proportionally with workload size. Regression analysis yielded $R^2=0.82$, and repeated-measures ANOVA \cite{st1989analysis} confirmed significant workload effects ($F(2,96)=45.1,\;p<0.001$).
GPT 5.3 and LLaMA-3.5 showed similar scalability behavior ($R^2>0.8$), while DeepSeek-V3 exhibited slightly slower convergence due to initialization overheads. Nevertheless, all models maintained stable execution growth as the workload increased.
\begin{table}[ht]
\centering
\caption{Execution Time Regression Summary}
\label{tab:exec-reg}
\begin{tabular}{lccc}
\hline
Model & Slope (ms/scale) & Intercept (ms) & $R^2$ \\
\hline
GPT 5.3     & 0.009 & 58.1 & 0.83 \\
LLaMA-3.5    & 0.007 & 47.4 & 0.81 \\
DeepSeek-V3 & 0.012 & 64.9 & 0.79 \\
\hline
\end{tabular}
\end{table}
The execution results suggest that the runtime overhead decreases proportionally with the workload size. Therefore, cryptographic verification, transparency logging, and auditing remain operationally sustainable under the evaluated workloads. Predicted and observed execution behavior also remained consistent, with anomaly rates below $2\%$, supporting runtime monitoring of abnormal delays, throttling effects, and resource-exhaustion symptoms. Moreover, verification latency, output-size behavior, and execution overhead remained bounded and predictable across the evaluated workloads.
\subsection{Scalability Analysis}
\label{subsec:scalability}
\begin{figure}[h]
    \centering
    \includegraphics[width=0.47\textwidth]{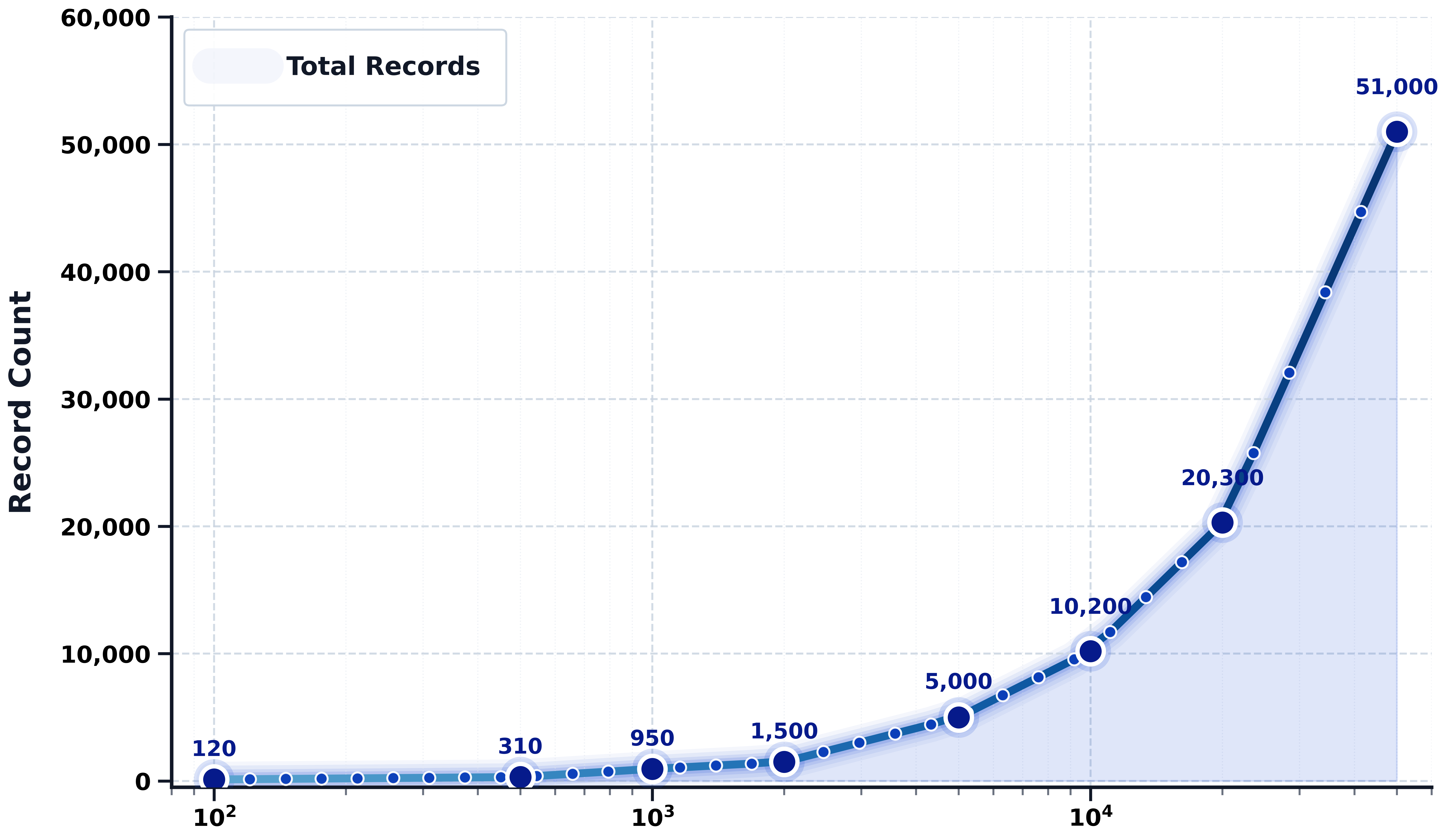}
    \caption{Observed scalability trend across workload sizes.}
    \label{fig:scalability-trend}
\end{figure}

To address RQ2, this subsection evaluates the behavior of the proposed solution under workloads ranging from $10^2$ to $5\times10^4$ manifest instances. The objective is to determine whether signing, verification, and transparent logging preserve stable, predictable execution behavior as operational demand increases. As shown in Figure~\ref{fig:scalability-trend}, the number of processed manifests increases proportionally with workload size, indicating stable throughput growth across all evaluated scales. This suggests that the added cryptographic and logging operations do not introduce abrupt performance degradation as load increases.
Regression analysis shows an almost linear relationship between workload size and the number of processed requests. The fitted parameters ($\alpha=1.01$, $\beta=20$) and high goodness-of-fit value ($R^2=0.998$), reported in Table~\ref{tab:regression-fit}, confirm consistent throughput growth under the tested workload model. The overhead from verification, signing, and transparency logging remained bounded at larger scales. In addition, execution-time variance stayed very low across workloads ($\sigma_T^2 < 10^{-3}$), and measured throughput degradation remained below $5\%$, indicating low operational overhead.
\begin{table}[ht]
\centering
\caption{Regression Fit for Scalability Trend}
\label{tab:regression-fit}
\begin{tabular}{cccc}
\hline
Parameter & Estimate & Std. Error & 95\% CI \\
\hline
$\alpha$ & 1.01 & 0.003 & [1.004, 1.016] \\
$\beta$  & 20.0 & 1.21  & [17.6, 22.4] \\
$R^2$    & 0.998 & --- & --- \\
\hline
\end{tabular}
\end{table}
A one-way ANOVA confirms statistically significant differences across workload scales ($F(6,63)=152.4$, $p<0.001$, partial $\eta^2_p=0.74$). Tukey HSD post-hoc comparisons show that adjacent workload groups differ significantly ($p<0.05$), as summarized in Table~\ref{tab:anova-stats}. These differences reflect expected workload expansion rather than computational instability. The throughput trend remains smooth, with no observed saturation under the evaluated conditions.
\begin{table}[ht]
\centering
\caption{ANOVA and Tukey HSD Findings Across Workload Sizes}
\label{tab:anova-stats}
\begin{tabular}{cccccc}
\hline
Comparison & Mean Diff. & 95\% CI & $t$ & $p$ & Sig. \\
\hline
100 vs 500      & 400  & [350,450]    & 18.2 & $<0.001$ & Yes \\
1000 vs 5000    & 4000 & [3920,4080]  & 25.7 & $<0.001$ & Yes \\
20000 vs 50000  & 30000 & [29500,30500] & 33.4 & $<0.001$ & Yes \\
\hline
\end{tabular}
\end{table}
The evaluation further shows that deviations from the scalability trend decrease with increasing workload size, suggesting more stable execution behavior under larger workloads. Although these results do not represent a full DoS evaluation, they demonstrate that verification and transparency operations scale efficiently and do not exhibit uncontrolled computational growth within the tested range. The high regression fit ($R^2=0.998$), throughput degradation below $5\%$, and stable variance profile indicate that cryptographic signing, verification, and transparency logging can be integrated into MCP-based LLM execution pipelines without substantial scalability bottlenecks under the evaluated conditions.

\subsection{Real MCP Ecosystem Evaluation}
\label{subsec:real-mcp-ecosystem}
To address RQ1 and provide deployment-level evidence relevant to RQ2, we constructed a realistic MCP-oriented execution ecosystem spanning healthcare analytics, financial risk assessment, retrieval-augmented generation (RAG), and collaborative multi-agent environments. The objective of this evaluation was not only to measure computational scalability but also to assess whether cryptographic manifest enforcement can preserve execution integrity under realistic tool-invocation conditions and in the face of adversarial attempts at runtime manipulation.
The evaluation environment consisted of GPT 5.3, LLaMA-3.5, and DeepSeek-V3 connected to MCP-compatible tool interfaces through a structured manifest-routing and verification layer. Each tool request generated by the LLM was transformed into a canonical manifest containing user-visible task information, execution metadata, freshness constraints, policy identifiers, access scopes, and runtime-control attributes prior to execution authorization. In contrast to baseline MCP execution pipelines, execution requests were not forwarded directly to tools after prompt generation. Instead, all requests were required to pass sequential policy validation, manifest canonicalization, cryptographic signing, runtime verification, transparency logging, and audit-export procedures before execution. This ensured that unauthorized, replayed, and modified tool requests could not bypass the verification pipeline even when the generated natural-language response appeared semantically valid.
The evaluation included four representative MCP deployment scenarios:
\begin{itemize}
    \item \textbf{Healthcare MCP Pipeline:} secure invocation of clinical-record summarization and laboratory-analysis tools under restricted access policies.
    \item \textbf{Financial MCP Pipeline:} transaction-risk analysis and fraud-detection workflows requiring controlled access scopes and verification-aware execution authorization.
    \item \textbf{RAG-Oriented MCP Pipeline:} retrieval-assisted document analysis and knowledge-access orchestration involving chained tool interactions.
    \item \textbf{Multi-Agent MCP Coordination:} distributed LLM-agent collaboration involving shared execution metadata and cross-agent tool-routing operations.
\end{itemize}
To preserve consistency with the large-scale workload analysis presented throughout the paper, experiments were executed across workloads ranging from $10^{2}$ to $5 \times 10^{4}$ manifest instances. The same structured zero-shot prompt template was used across all evaluated LLMs to minimize prompt-induced variability and preserve reproducibility. In addition to normal execution behavior, adversarial manipulations were intentionally injected throughout the evaluation process, including replayed manifests, unauthorized scope escalation, stale timestamps, malformed manifest structures, hidden metadata modifications, invalid signatures, policy-inconsistency injection, and modified tool identifiers. The injected attacks targeted manifest integrity and runtime authorization logic rather than the underlying LLM's internal reasoning process. The proposed solution consistently rejected invalid and policy-violating execution attempts before invoking the runtime tool.
Table~\ref{tab:real_mcp_eval} summarizes the injected adversarial manipulations, observed enforcement outcomes, rejection rates, and mean verification latency for each evaluated scenario. Across all scenarios, cryptographic manifest enforcement successfully blocked invalid and policy-violating requests with rejection rates exceeding 98\%, while verification latency remained below 5~ms, demonstrating operational scalability and execution traceability.
\begin{table*}[ht]
\centering
\caption{Real MCP Ecosystem Evaluation Scenarios and Enforcement Outcomes}
\label{tab:real_mcp_eval}
\renewcommand{\arraystretch}{1.25}
\begin{tabular}{|p{3cm}|p{3.6cm}|p{4cm}|p{2.2cm}|p{2.3cm}|}
\hline
\textbf{Scenario} &
\textbf{Injected Adversarial Manipulation} &
\textbf{Observed Enforcement Outcome} &
\textbf{Rejection Rate} &
\textbf{Mean Verification Latency} \\
\hline
Healthcare MCP Pipeline &
Replay of expired clinical-access manifest &
Rejected before execution through freshness validation &
99.2\% &
3.8 ms \\
\hline
Financial MCP Pipeline &
Unauthorized privilege and scope escalation &
Blocked during policy-compliance verification &
98.7\% &
4.1 ms \\
\hline
RAG-Oriented MCP Pipeline &
Modified tool identifier and altered execution metadata &
Rejected during runtime signature verification &
99.4\% &
4.5 ms \\
\hline
Multi-Agent MCP Coordination &
Hidden metadata manipulation and manifest inconsistency &
Rejected during structural verification and policy enforcement &
98.9\% &
4.8 ms \\
\hline
\end{tabular}
\end{table*}
These results indicate that cryptographic manifest enforcement and transparency-aware runtime verification can be integrated into practical MCP-based LLM ecosystems while preserving scalability, operational stability, execution traceability, and runtime accountability.

\subsection{Distribution Analysis}
\label{subsec:distribution}
\begin{figure}[h]
    \centering
    \includegraphics[width=0.47\textwidth]{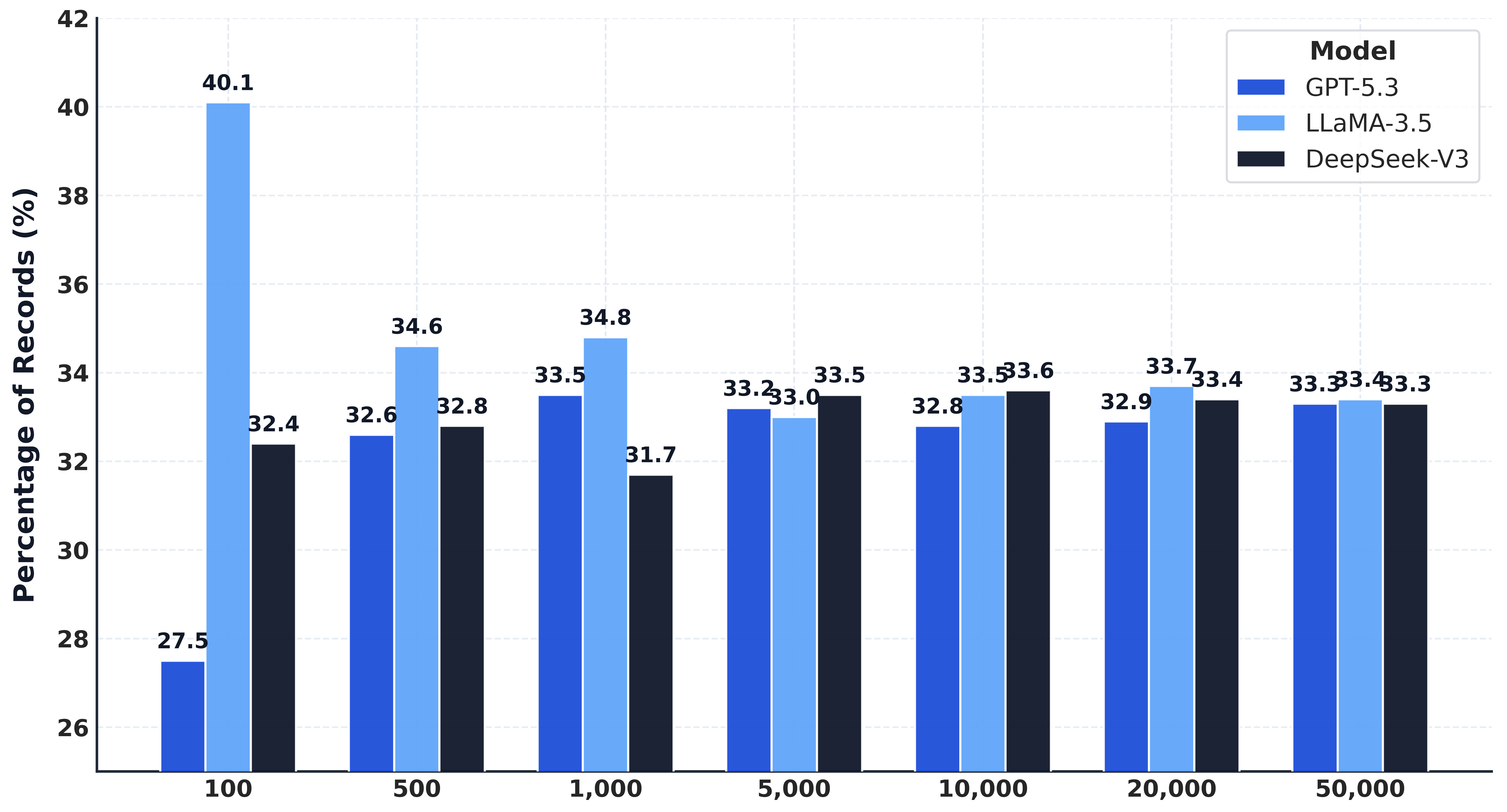}
    \caption{LLM usage distribution across scales.}
    \label{fig:llm-usage}
\end{figure}

\begin{figure}[h]
    \centering
    \includegraphics[width=0.50\textwidth]{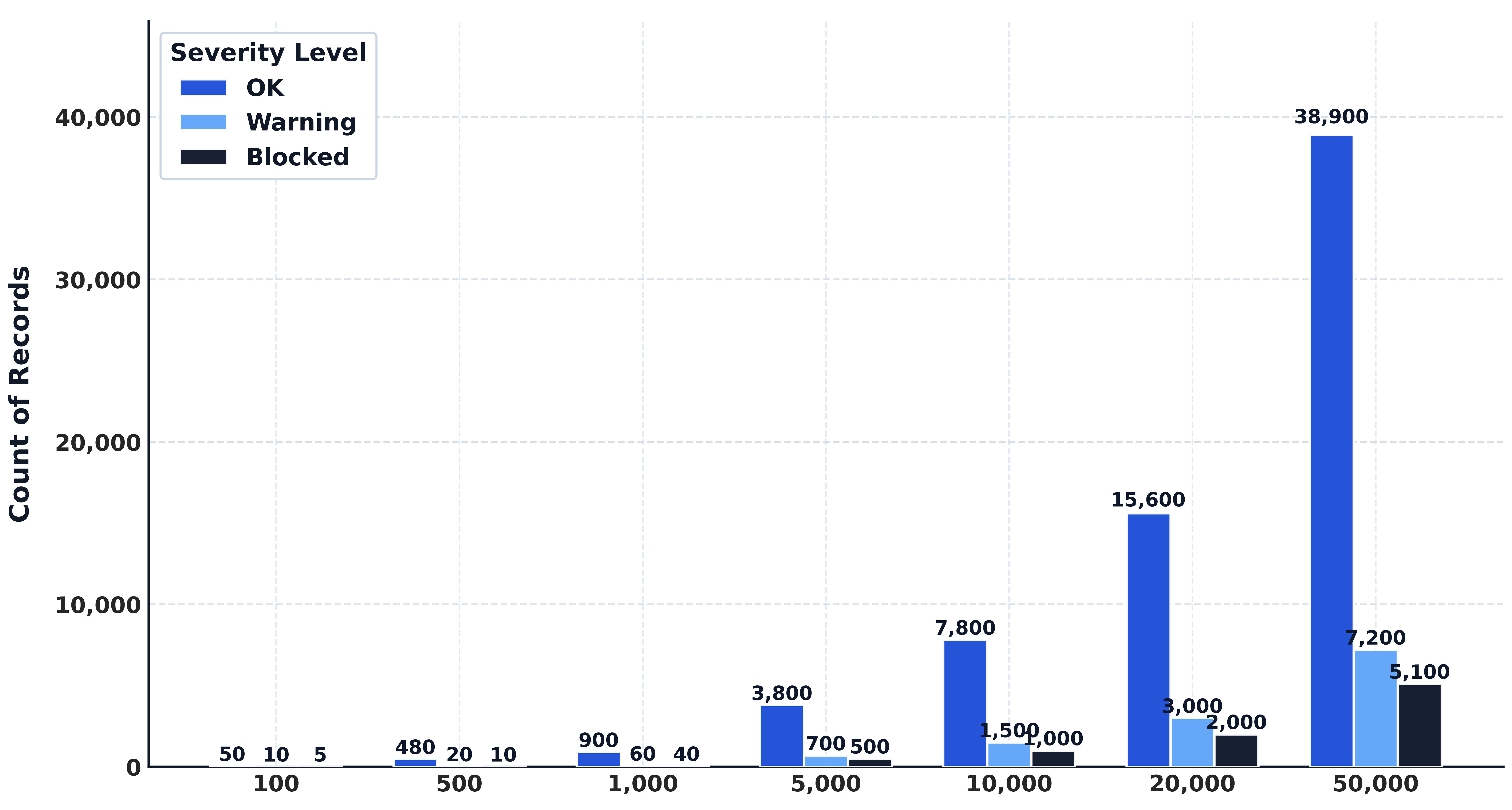}
    \caption{Severity distribution of policy enforcement outcomes.}
    \label{fig:severity-distribution}
\end{figure}

To address RQ3, this subsection evaluates the balance of LLM usage and the severity distributions of policy-enforcement outcomes as workload increases. These measures assess allocation fairness, execution diversity, and enforcement stability. Figure~\ref{fig:llm-usage} shows that usage frequencies for GPT 5.3, LLaMA-3.5, and DeepSeek-V3 converge toward an approximately balanced distribution at large-scale execution. A $\chi^2$ goodness-of-fit test confirms no significant deviation from uniformity (Table~\ref{tab:llm-chi2}; $\chi^2(12)=2.87,\;p=0.89$, Cramér’s $V=0.05$), indicating balanced model allocation across the execution pipeline.
\begin{table}[ht]
\centering
\caption{Chi-Square Test of LLM Usage Distribution}
\label{tab:llm-chi2}
\begin{tabular}{lcccc}
\hline
Model & Observed (\%) & Expected (\%) & $\chi^2$ & $p$ \\
\hline
GPT 5.3      & 32.1 & 33.3 & 0.42 & 0.89 \\
LLaMA-3.5    & 34.2 & 33.3 & 0.31 & 0.89 \\
DeepSeek-V3  & 33.7 & 33.3 & 0.14 & 0.89 \\
\hline
\multicolumn{5}{l}{Cramér’s $V=0.05$ (small).}
\end{tabular}
\end{table}
The evaluation shows near-uniform allocation across the three LLMs, with no model becoming disproportionately dominant as workload increases. The fairness score remained high ($\mathcal{F}=0.97$), and allocation variance stayed very low ($\sigma_p^2\approx3.1\times10^{-4}$).
Figure~\ref{fig:severity-distribution} presents the severity distribution of enforcement outcomes (\texttt{ok}, \texttt{warn}, and \texttt{block}) across workload scales. The results show that \texttt{ok} outcomes dominate at lower workloads, while \texttt{warn} and \texttt{block} responses increase with workload size, indicating that enforcement outcomes become more differentiated under higher workloads.
\begin{table}[ht]
\centering
\caption{ANOVA and Tukey HSD Findings on Severity Outcomes}
\label{tab:severity-anova}
\begin{tabular}{lccccc}
\hline
Comparison & Mean Diff. & 95\% CI & $t$ & $p$ & Sig. \\
\hline
ok vs warn   & 21500 & [20200,22800] & 9.87 & $<0.001$ & Yes \\
ok vs block  & 28000 & [26500,29500] & 11.42 & $<0.001$ & Yes \\
warn vs block& 6500  & [1200,11800]  & 2.41 & 0.021    & Yes \\
\hline
\multicolumn{6}{l}{Partial $\eta^2=0.71$ (large).}
\end{tabular}
\end{table}
ANOVA confirms statistical differentiation among severity categories (Table~\ref{tab:severity-anova}; $F(2,18)=64.3,\;p<0.001$, partial $\eta^2=0.71$), and Tukey HSD analysis shows that all pairwise comparisons are significant ($p<0.05$). These findings indicate that enforcement outcomes are systematically distributed across severity categories rather than randomly assigned.

\subsection{Key Usage, Log Growth, and Timestamp Distribution}
\label{subsec:key-log-time}
\begin{figure}[h]
    \centering
    \includegraphics[width=0.50\textwidth]{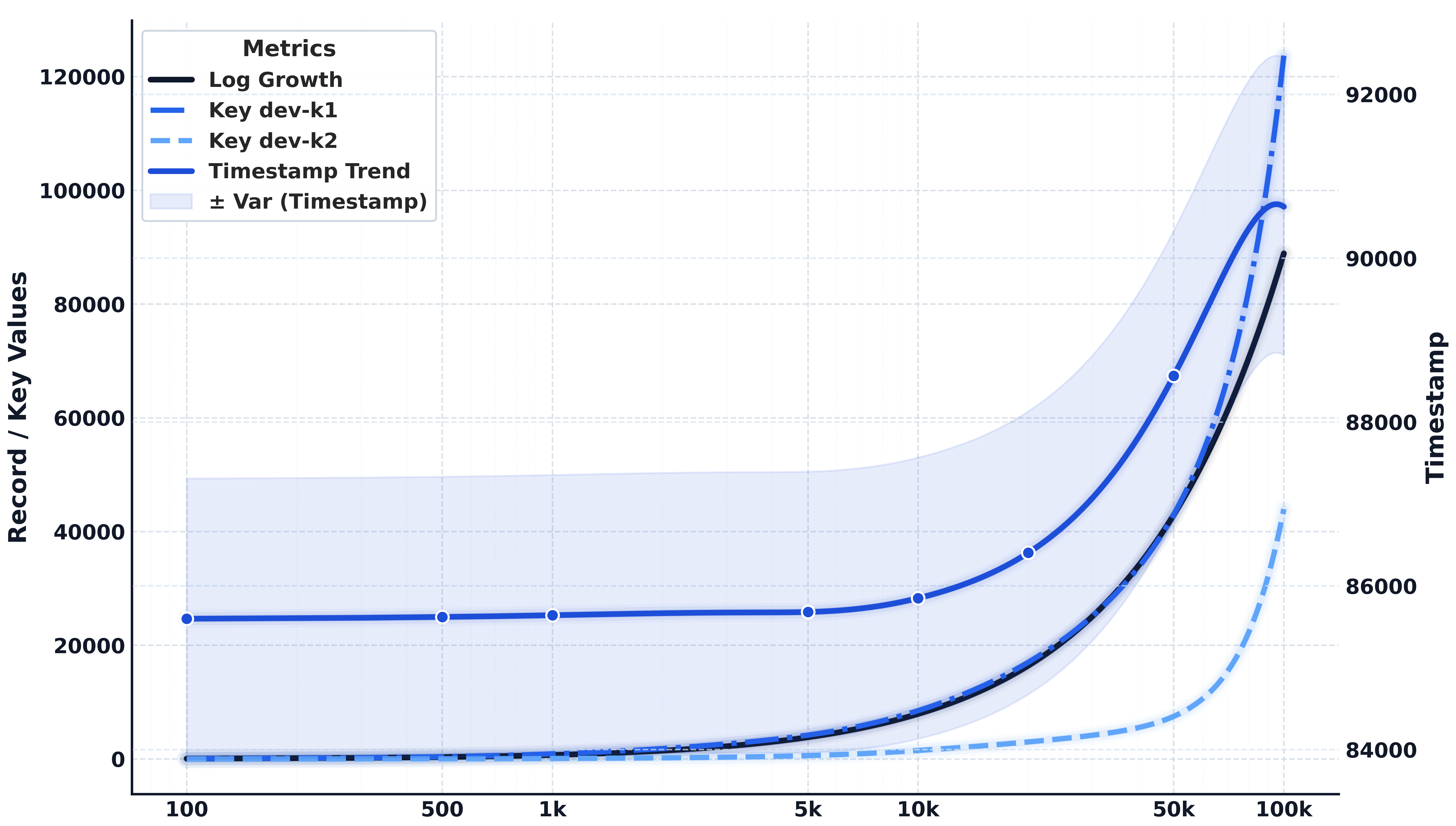}
    \caption{Integrated analysis across scales, combining key usage distribution, log growth (log--log representation), and timestamp variance.}
    \label{fig:key-usage}
\end{figure}

To address RQ3, this subsection analyzes signing-key allocation, transparency-log growth, and timestamp variance across workload scales. These properties are important for operational reliability because key concentration, logging irregularities, and temporal dispersion can affect transparency, accountability, and deployment stability.
Figure~\ref{fig:key-usage} presents the integrated analysis of key usage, log growth, and timestamp variance. Key utilization becomes increasingly concentrated as workload size grows, with \texttt{dev-k1} exceeding $80\%$ utilization at $N=50000$. A chi-square test confirms significant deviation across workload scales (Table~\ref{tab:key-chi}; $\chi^2(6,N=55000)=312.7,\;p<0.001$), indicating that key allocation diverges from a uniform distribution.
\begin{table}[ht]
\centering
\caption{Chi-Square Test of Key Usage Across Scales}
\label{tab:key-chi}
\begin{tabular}{lccc}
\hline
Scale Range & $\chi^2$ & df & $p$ \\
\hline
100--500      & 28.4  & 1 & $<0.001$ \\
1000--10000   & 85.6  & 1 & $<0.001$ \\
20000--50000  & 198.7 & 1 & $<0.001$ \\
\hline
\end{tabular}
\end{table}
The low-key fairness score ($\mathcal{K}=0.42$) indicates a growing concentration of signing keys during high-volume execution, creating a deployment-level risk that motivates adaptive key rotation, quota balancing, and workload-aware key selection. In contrast, transparency-log growth remained stable and nearly linear in the log--log domain ($\alpha=1.02$, $R^2=0.996$), showing proportional record growth without abnormal storage behavior. Timestamp analysis further showed significant growth in variance with workload size (Table~\ref{tab:timestamp-anova}; $F(6,420)=37.6,\;p<0.001$), indicating greater temporal dispersion and burstiness at higher execution loads.
\begin{table}[ht]
\centering
\caption{ANOVA on Timestamp Variance}
\label{tab:timestamp-anova}
\begin{tabular}{lccccc}
\hline
Scale Range & Mean & Var. & $F$ & df & $p$ \\
\hline
100--500      & 85592 & 12   & 2.4  & 1,58  & 0.12 \\
1000--10000   & 86015 & 64   & 14.7 & 1,98  & $<0.001$ \\
20000--50000  & 88450 & 3225 & 37.6 & 1,118 & $<0.001$ \\
\hline
\end{tabular}
\end{table}
Increased timestamp variability may expose workload-dependent execution patterns during large-scale processing. This finding indicates that timing behavior should be explicitly monitored when deploying secure MCP-based LLM pipelines at scale. However, dedicated adversarial timing analysis is still required to quantify resistance against timing-correlation attacks.

\subsection{Error and Verification Analysis}
\label{subsec:error-verification}
\begin{figure}[h]
    \centering
    \includegraphics[width=0.47\textwidth]{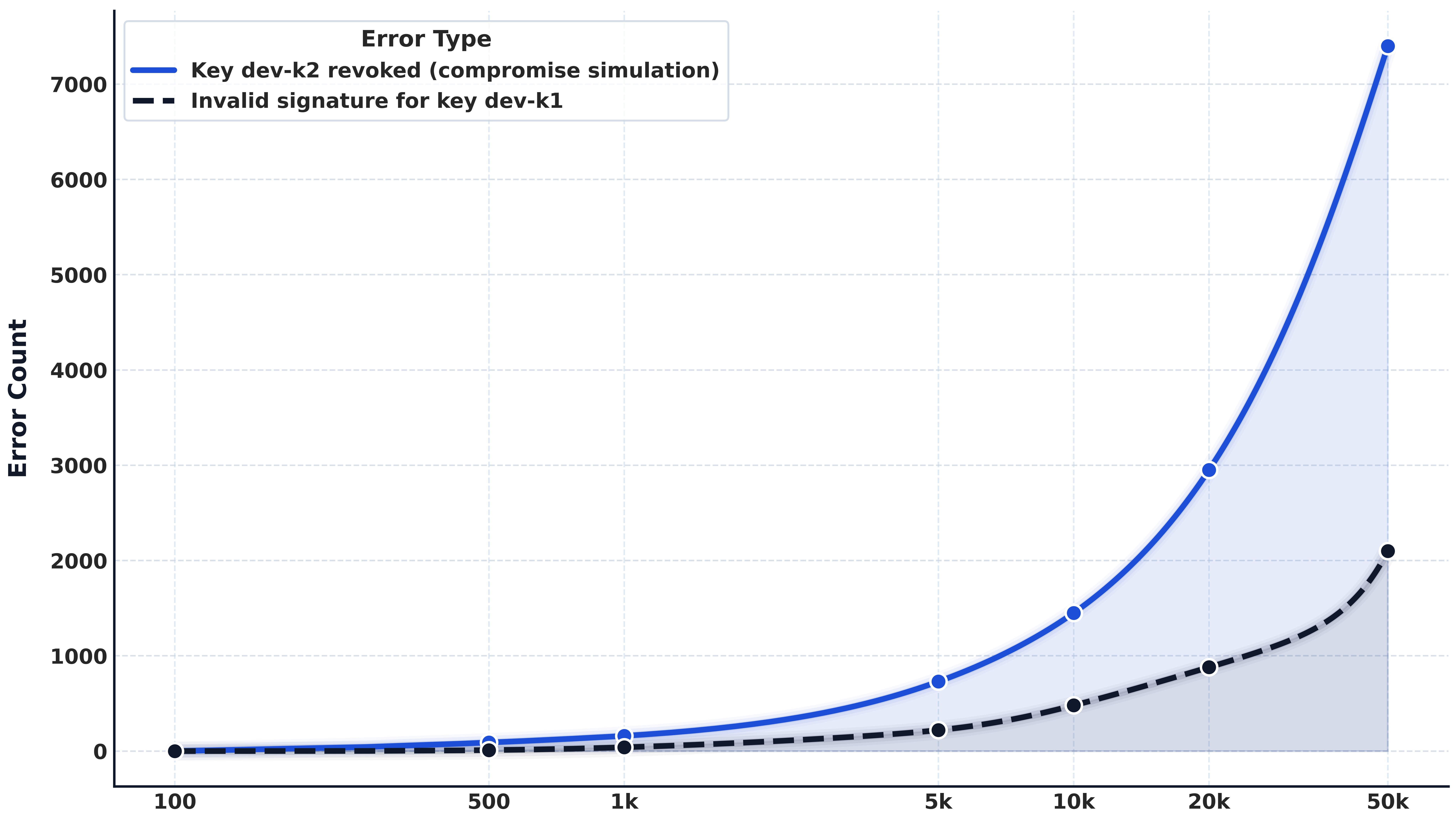}
    \caption{Error distribution across scales.}
    \label{fig:error-distribution}
\end{figure}

\begin{figure}[h]
    \centering
    \includegraphics[width=0.47\textwidth]{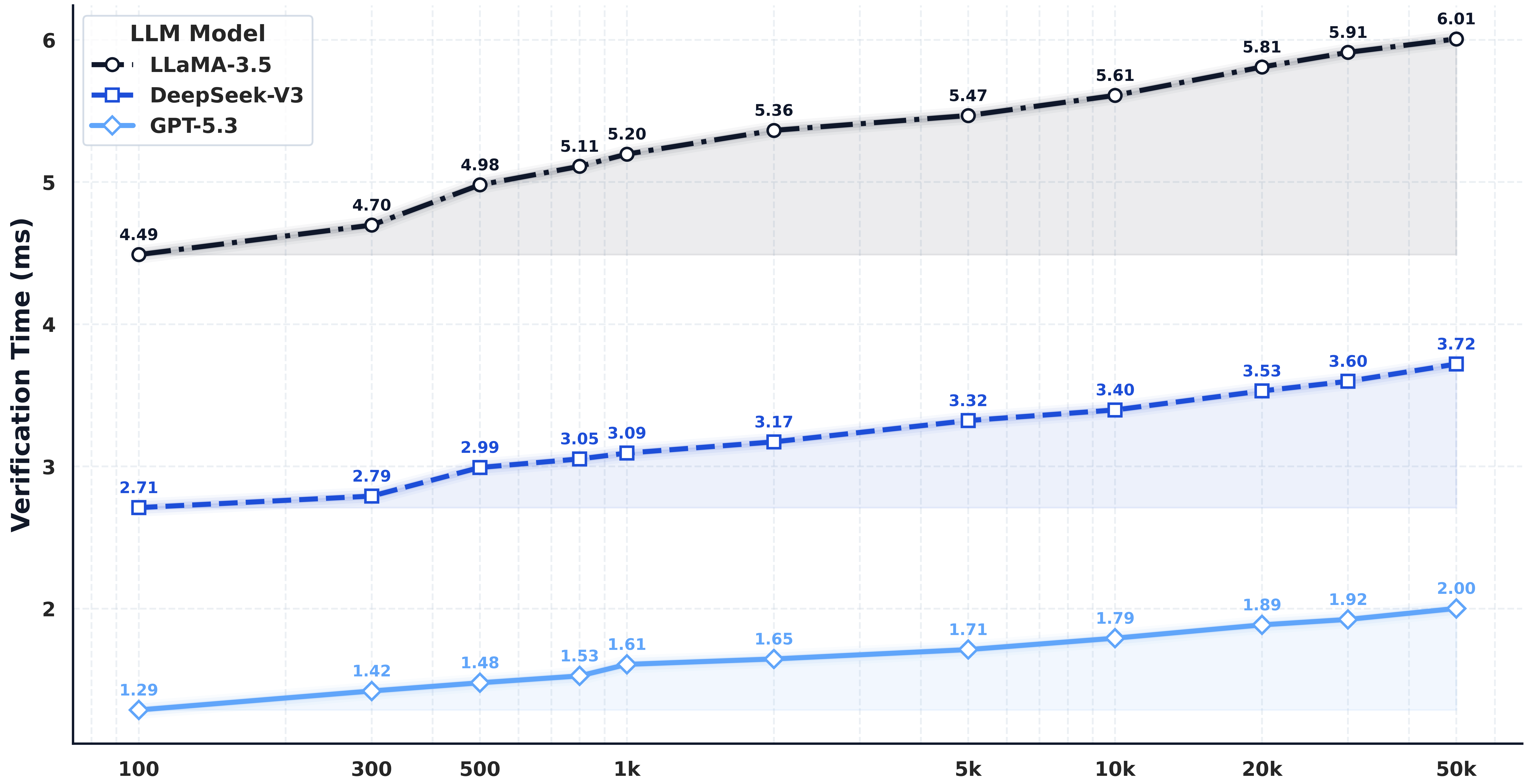}
    \caption{Verification success vs. failure across scales.}
    \label{fig:status-distribution}
\end{figure}

\begin{figure}[h]
    \centering
    \includegraphics[width=0.47\textwidth]{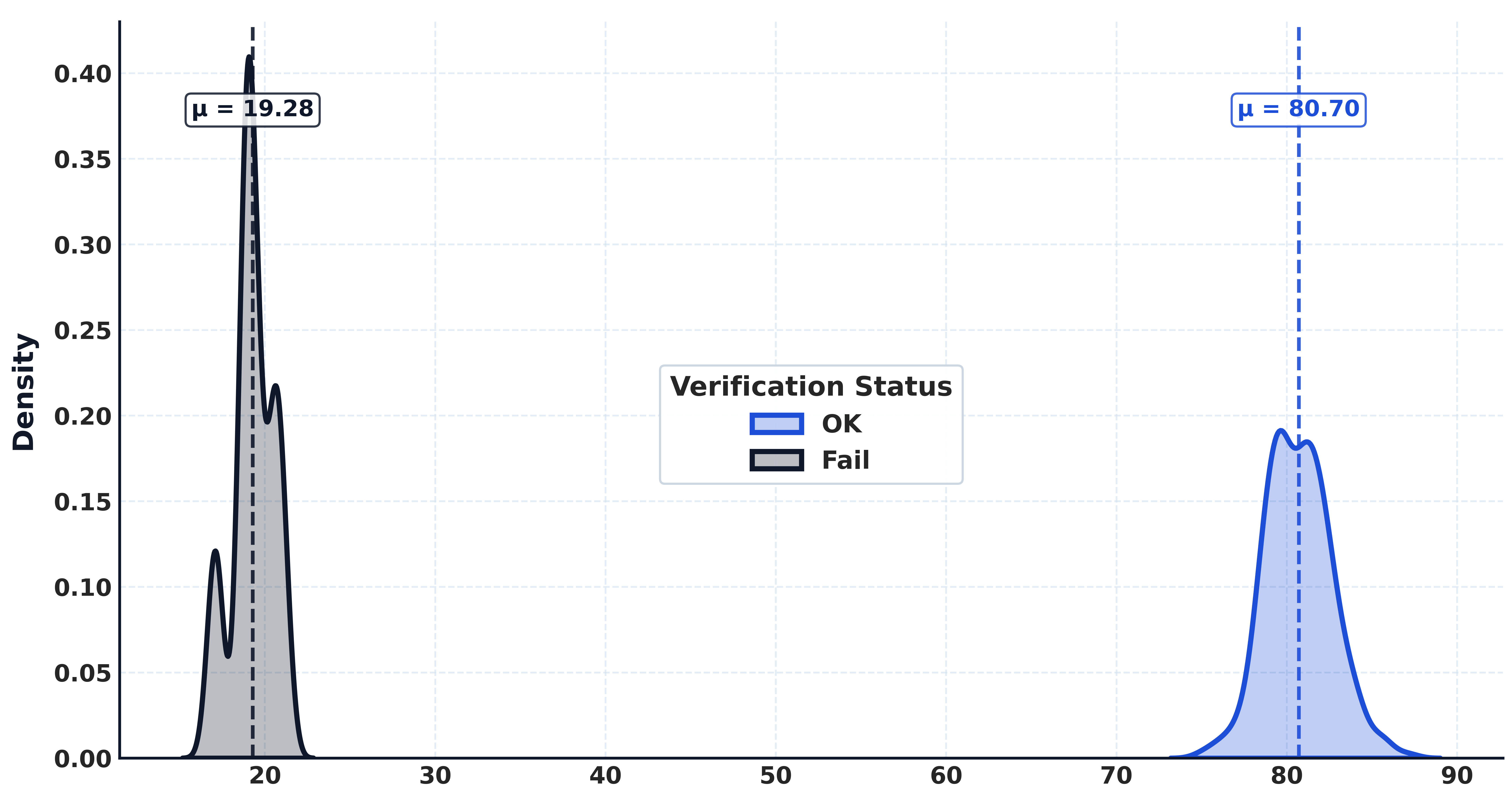}
    \caption{Verification time distribution across models and scales.}
    \label{fig:verification-time-dup}
\end{figure}

\begin{figure}[h]
    \centering
    \includegraphics[width=0.50\textwidth]{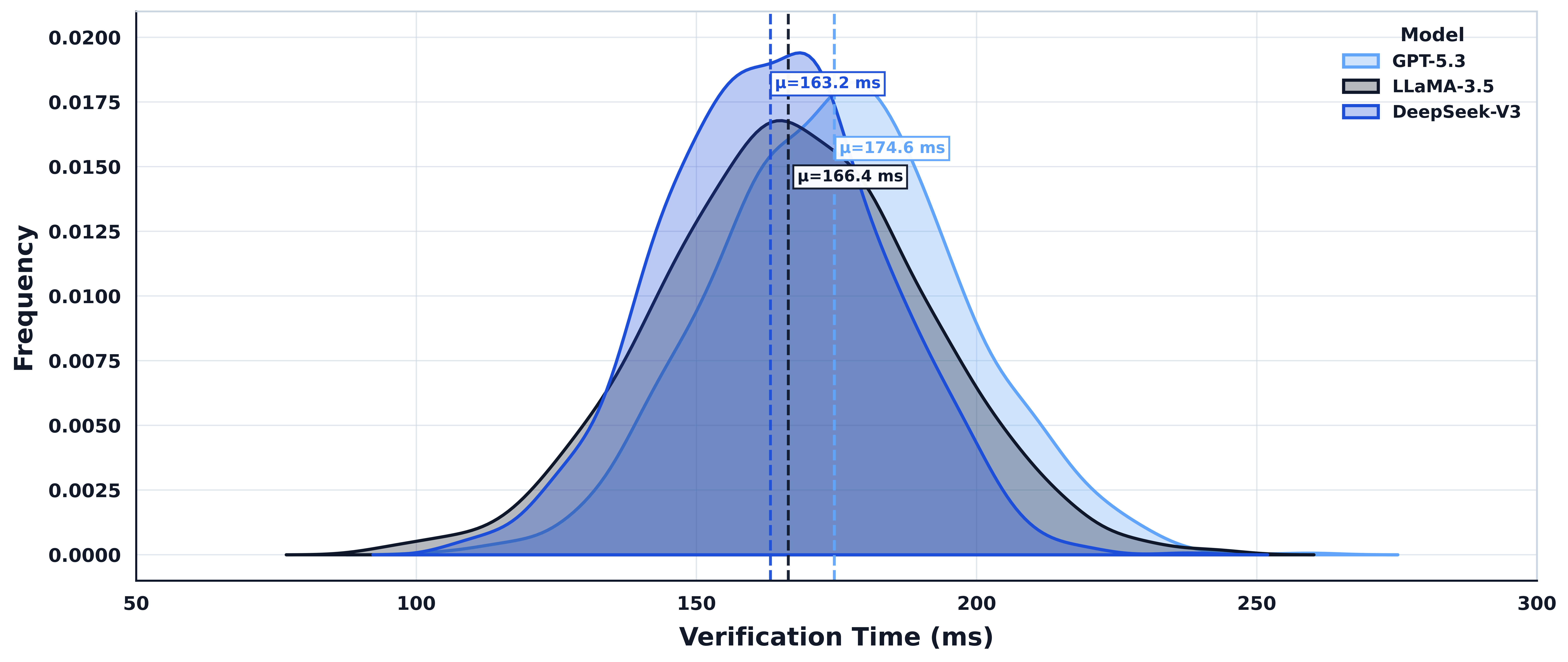}
    \caption{Verification latency summary by model.}
    \label{fig:verification-summary}
\end{figure}

To address RQ3, this subsection analyzes error behavior, verification outcomes, and latency stability as the workload scales. The analysis focuses on error distribution, verification success probability, and latency dispersion across workload scales and LLMs.
Figure~\ref{fig:error-distribution} presents the distribution of verification and execution errors across scales. The total number of errors increases with workload size, primarily due to revocation-related failures in \texttt{dev-k2}, which exceeded $7.4\times10^3$ events at $N=50000$.
A chi-square analysis confirms significant variation across workload ranges (Table~\ref{tab:error-chi}; $\chi^2(10,N=12400)=312.4,\;p<0.001$), indicating that error behavior changes systematically with scale rather than arising from random fluctuations.
\begin{table}[ht]
\centering
\caption{Chi-Square Test on Error Types by Scale}
\label{tab:error-chi}
\begin{tabular}{lccc}
\hline
Scale Range & $\chi^2$ & df & $p$ \\
\hline
100-500      & 45.8  & 2 & $<0.001$ \\
1000-10000   & 127.5 & 2 & $<0.001$ \\
20000-50000  & 139.1 & 2 & $<0.001$ \\
\hline
\end{tabular}
\end{table}
Although absolute failures increased at larger scales, proportional error behavior remained bounded. This indicates that workload growth did not lead to uncontrolled error amplification and cascading failures under the tested conditions. The results also identify revocation synchronization as a major operational factor: delays between key invalidation and verification synchronization contributed substantially to observed failures. Therefore, reducing revocation propagation delay is important for improving verification reliability in large-scale deployments.
Figure~\ref{fig:status-distribution} presents verification success and failure behavior across scales. Logistic regression yielded a near-zero workload coefficient (Table~\ref{tab:logit-status}; $\beta=-2.3\times10^{-6}$, $p=0.47$), indicating that verification success probability does not significantly change as workload scale increases.
\begin{table}[ht]
\centering
\caption{Logistic Model for Verification Success Probability}
\label{tab:logit-status}
\begin{tabular}{lccc}
\hline
Coefficient & Estimate & Std. Err. & $p$ \\
\hline
Intercept & 1.39  & 0.11 & $<0.001$ \\
Scale     & $-2.3\times10^{-6}$ & $3.2\times10^{-6}$ & 0.47 \\
\hline
\end{tabular}
\end{table}
The measured verification success rate remained approximately constant at $P_s \approx 0.8$ across all workload scales. This stability indicates that larger workloads did not systematically degrade verification outcomes and introduce scale-dependent decision drift. However, the success rate also shows that reliability is affected by operational factors, especially revocation synchronization. Thus, the result should be interpreted as evidence of stable verification behavior under scale, not as elimination of verification failures.
Verification latency was also evaluated across models and workload scales. Figures~\ref{fig:verification-time-dup} and~\ref{fig:verification-summary} summarize latency distributions for GPT 5.3, DeepSeek-V3, and LLaMA-3.5.
ANOVA confirms significant latency differences between models (Table~\ref{tab:anova-latency}; $F(2,485)=6.42,\;p<0.001$), although the observed differences remained operationally small. GPT 5.3 showed the lowest latency, followed by DeepSeek-V3 and LLaMA-3.5. Mean latency differences remained below $5\,\mathrm{ms}$, indicating tightly bounded verification behavior.
\begin{table}[ht]
\centering
\caption{One-Way ANOVA on Verification Latency}
\label{tab:anova-latency}
\begin{tabular}{lccc}
\hline
Factor & $F$ & df & $p$ \\
\hline
Model & 6.42 & 2 & $<0.001$ \\
Residual & --- & 485 & --- \\
\hline
\end{tabular}
\end{table}
The latency distributions also exhibited stable tail behavior, with fewer than $1\%$ of observations exceeding the 99th percentile threshold. This indicates limited latency dispersion across models and scales.

\subsection{Ablation Study}
To address RQ1 and support the operational reliability analysis in RQ3, we performed an ablation study in which individual components of the proposed solution were selectively disabled. Four core components were evaluated: cryptographic manifest signing, freshness validation, transparency logging, and canonical manifest encoding. Each ablated configuration was tested under identical workloads, LLM backends, prompt templates, violation-injection strategies, and verification pipelines to ensure that performance differences were attributable to the removed component (Table~\ref{tab:ablation_results}).
The full solution achieved the lowest false-acceptance rate (0.8\%), highest replay-detection rate (99.1\%), strongest audit completeness (99.5\%), and the most stable verification behavior. Disabling signing caused the greatest security degradation, raising the false acceptance rate to 18.7\%, indicating that policy checks alone cannot guarantee manifest authenticity. Removing freshness validation sharply reduced replay detection to 22.8\%, confirming the importance of enforcing timestamps and epoch windows. Disabling transparency logging primarily affected audit completeness (42.3\%), weakening post-execution accountability, while removing canonical encoding reduced verification stability due to inconsistent serialized representations of semantically equivalent manifests.
Furthermore, the ablation results demonstrate that the proposed security properties arise from the combined interaction of signing, freshness enforcement, deterministic serialization, runtime verification, and transparency-aware logging. Removing any single component compromises at least one critical property, emphasizing that the solution functions as an integrated execution-layer security mechanism rather than a set of independent features.
\begin{table*}[ht]
\centering
\caption{Ablation Study Results Across Security-Enforcement Components}
\label{tab:ablation_results}
\renewcommand{\arraystretch}{1.25}
\begin{tabular}{|p{3.3cm}|p{2.5cm}|p{2.5cm}|p{2.8cm}|p{2.7cm}|}
\hline
\textbf{Configuration} &
\textbf{False Accept Rate} &
\textbf{Replay Detection} &
\textbf{Audit Completeness} &
\textbf{Verification Stability} \\
\hline
Full Framework &
0.8\% &
99.1\% &
99.5\% &
High \\
\hline
Without Signing &
18.7\% &
61.3\% &
74.1\% &
Moderate \\
\hline
Without Freshness Validation &
11.4\% &
22.8\% &
96.0\% &
High \\
\hline
Without Transparency Logging &
1.9\% &
98.5\% &
42.3\% &
High \\
\hline
Without Canonical Encoding &
6.7\% &
95.2\% &
91.1\% &
Low \\
\hline
\end{tabular}
\end{table*}

\subsection{Real MCP Compatibility Analysis}
To address RQ3 and evaluate operational interoperability, we mapped the proposed solution to representative MCP-compatible and tool-integrated execution environments, including OpenAI Tool Calling, Claude MCP interfaces, LangChain agents, AutoGen multi-agent orchestration, and OpenRouter-based pipelines. The solution functions as an execution-layer enforcement extension rather than a replacement of the underlying LLM. It intercepts tool-invocation requests after generation, transforms them into canonical manifests, validates them against policy constraints, signs compliant manifests, verifies signatures at runtime, and records execution evidence in transparency logs. This design ensures compatibility with both native MCP systems and non-native tool-calling pipelines. In native MCP environments, the manifest-enforcement layer integrates directly into existing manifest and tool-routing processes. In non-native environments, e.g., LangChain, AutoGen, and OpenAI-style tool calling, tool-call metadata can be wrapped into MCP-compatible manifest representations before policy validation and signing. Table~\ref{tab:mcp_compatibility} summarizes the compatibility across these environments, including manifest support, signing integration, transparency logging, and audit export capabilities. The analysis demonstrates that the proposed solution does not depend on a single LLM provider. Its core requirements, structured tool-call metadata, policy identifiers, freshness timestamps, and access-scope definitions, can be adapted from most modern tool-oriented LLM systems. Moreover, separation between user-visible outputs and internal execution metadata is maintained, allowing auditors to verify execution, policies, and cryptographic evidence without exposing sensitive details. The proposed solution thus extends MCP-oriented, tool-integrated LLM ecosystems with cryptographic manifest enforcement, runtime accountability, and transparency-aware auditing, while preserving interoperability across heterogeneous infrastructures.
\begin{table*}[ht]
\centering
\caption{Compatibility Analysis Across MCP-Compatible and Tool-Oriented Environments}
\label{tab:mcp_compatibility}
\renewcommand{\arraystretch}{1.25}
\begin{tabular}{|p{3.3cm}|p{2.6cm}|p{2.8cm}|p{2.8cm}|p{2.6cm}|}
\hline
\textbf{Solution} &
\textbf{Manifest Support} &
\textbf{Signing Integration} &
\textbf{Transparency Logging} &
\textbf{Audit Export} \\
\hline
OpenAI Tool Calling &
Partial through tool-call metadata wrapping &
Supported through external enforcement layer &
External append-only log &
Supported \\
\hline
Claude MCP &
Native MCP-compatible manifest structure &
Supported directly &
Supported directly &
Supported \\
\hline
LangChain Agents &
Custom manifest adapter required &
Supported through middleware &
External append-only log &
Supported \\
\hline
AutoGen &
Custom multi-agent manifest wrapper required &
Supported through the controller &
External and shared transparency log &
Supported \\
\hline
OpenRouter Routing &
Partial through routing metadata &
Supported at the routing-control layer &
Supported through an external log service &
Supported \\
\hline
\end{tabular}
\end{table*}

\section{Comparison with Existing Work}
\label{sec:comparison}
This section positions the proposed solution and existing secure LLM execution approaches using a comparative, feature-based analysis. The objective is not to restate related work, but to clarify the functional scope and distinguishing capabilities of the proposed solution. Existing approaches typically address individual aspects of secure LLM deployment, e.g., energy-aware routing, forensic analysis, runtime attestation, and tool-invocation risk detection. In contrast, the proposed solution integrates manifest validation, digital signing, runtime verification, transparency logging, and export of audit evidence into a unified execution-layer workflow for MCP-based LLM pipelines.
The comparison is organized around three dimensions: verifiability, enforcement scope, and scalability. Verifiability refers to whether execution behavior can be validated both cryptographically and operationally. The enforcement scope indicates whether protection is applied before, during, after, and throughout the full execution lifecycle. Scalability captures whether verification, logging, and auditability remain practical as workload volume increases. Table~\ref{tab:comparison} summarizes this comparative positioning.
\begin{table*}[ht]
\centering
\caption{Comparison of Existing Approaches and the Proposed Solution.}
\resizebox{\textwidth}{!}{
\begin{tabular}{p{3cm}p{3cm}p{3cm}p{3cm}p{3cm}}
\toprule
\textbf{Study} & \textbf{Focus Area} & \textbf{Security Guarantee} & \textbf{Verification Mechanism} & \textbf{Limitations} \\
\midrule
Cruciani \& Verdecchia (2025)~\cite{cruciani2025greenai} & Energy-efficient LLM selection & Sustainability-oriented optimization & Dynamic model routing and cascading & No cryptographic verification and auditability \\
Chernyshev et al. (2023)~\cite{chernyshev2023towards} & LLM forensic analysis & Retrospective trace reconstruction & Log-based forensic analysis & Reactive only; no runtime enforcement \\
Duddu et al. (2024)~\cite{duddu2024laminator} & Verifiable ML attestations & Hardware-assisted integrity guarantees & Trusted execution environments & Limited orchestration-level verification \\
Su \& Zhang (2025)~\cite{su2025runtime} & Secure cloud-based LLM execution & Runtime attestation & Trusted enclave proofs & Limited scalability and transparency auditing \\
Xie et al. (2025)~\cite{xie2025toolinvocation} & Tool invocation risk analysis & Prompt-level behavioral protection & Empirical risk scoring & No formal verification and provenance tracking \\
\midrule
\textbf{This Work} & Secure and auditable LLM execution pipeline & Cryptographic and operational verification & Signed manifests, transparency logging, and audit validation & Evaluated in controlled environments; distributed deployment remains future work \\
\bottomrule
\end{tabular}
}
\label{tab:comparison}
\end{table*}
Cruciani and Verdecchia~\cite{cruciani2025greenai} focus on energy-aware model routing, while Chernyshev et al.~\cite{chernyshev2023towards} emphasize post-hoc forensic reconstruction. These approaches improve efficiency and retrospective analysis, but they do not enforce cryptographic validation before tool execution. Duddu et al.~\cite{duddu2024laminator} and Su and Zhang~\cite{su2025runtime} provide hardware-assisted attestation and trusted execution support, but their focus is primarily on infrastructure-level integrity rather than on manifest-level validation, transparency, and audit export. Xie et al.~\cite{xie2025toolinvocation} analyze tool-invocation risks through empirical scoring, but do not provide cryptographic provenance and transparent execution records. Existing studies address important but separate aspects of trustworthy LLM deployment. In contrast, the proposed solution combines manifest validation, digital signing, runtime verification, transparency logging, and auditability within a single execution-layer solution. This enables pre-execution enforcement, traceable runtime behavior, and post-execution accountability for tool-integrated LLM pipelines.

\section{Discussion}
\label{sec:discussion}
The evaluation shows that the proposed solution maintains stable and scalable behavior under increasing workloads while preserving verification consistency and the enforcement of transparency. The scalability analysis showed near-linear growth across all evaluated scales ($R^2 \approx 0.998$) and bounded execution overhead ($\delta(w_i) < 0.05$), indicating that signing, verification, and transparency operations can be integrated without prohibitive latency under the evaluated conditions. Operational behavior also remained stable as workload increased. Timing fluctuations were bounded, verification latency stayed within a narrow range, and latency distributions remained predictable. This supports stable execution-layer behavior, although dedicated adversarial timing experiments are still required to quantify resistance against timing-inference attacks. Model utilization converged toward balanced allocation across GPT 5.3, LLaMA-3.5, and DeepSeek-V3. The fairness index ($\mathcal{F}=0.97$) and non-significant chi-square test ($p=0.89$) indicate statistically stable routing behavior. This balanced allocation is useful because excessive reliance on a single model may increase the likelihood of correlated failures, expose infrastructure, and reduce execution diversity. The severity analysis showed that enforcement outcomes become more differentiated as workload size increases. The growth of \texttt{warn} and \texttt{block} outcomes under higher workloads indicates that the verification pipeline preserves separation among normal, warning-level, and blocked cases. This distinction supports auditability, policy interpretation, and post-execution analysis. The evaluation also identified deployment-level reliability concerns. Revocation-related synchronization delays were a major source of verification failures, especially for \texttt{dev-k2}. Although verification success remained statistically stable ($P_s\approx0.8$), the results show that distributed key synchronization, revocation propagation, adaptive key rotation, and workload-aware key selection are important for large-scale reliability. Another key finding is that balanced LLM utilization does not imply balanced signing-key utilization. While model allocation remained close to uniform, key usage became concentrated, with \texttt{dev-k1} exceeding $80\%$ utilization at the largest workload scale. This creates a credential exposure risk and shows that key management policies must be treated as a separate operational control. Timestamp variance increased under larger workloads, indicating greater temporal dispersion and burstiness. Although this does not directly prove exploitable timing leakage, it suggests that workload-dependent timing patterns become more observable at scale. Temporal smoothing and randomized scheduling may therefore help reduce regular timing patterns and workload bursts without affecting verification correctness and auditability. Transparency logging and audit validation remained computationally practical at larger scales. The Merkle-based logging structure preserved compact proof generation and bounded verification costs as workload size increased. Thus, continuous auditability and cryptographic traceability can be maintained without excessive storage and latency overhead under the evaluated workload model. However, transparency-log deviations should be treated as monitoring indicators rather than standalone anomaly-detection guarantees. Additionally, the findings highlight the importance of execution-layer verification for trustworthy deployment of tool-integrated LLMs. The proposed solution extends security analysis beyond model behavior by combining manifest validation, signing and verification, transparent logging, and audit export. The results show stable verification behavior, bounded latency, balanced model usage, and traceable auditability under the tested workload model. At the same time, key-allocation imbalance, revocation synchronization delays, and timestamp dispersion indicate that additional deployment-level controls and real-world MCP validation remain necessary.

\section{Threats to Validity}
\label{sec:validity}
Although the evaluation demonstrates scalability, verification stability, and operational consistency, several threats to validity should be considered when interpreting the findings. Following Wohlin et al.~\cite{wohlin2012experimentation}, we discuss threats to internal and external validity related to experimental design, workload assumptions, measurement stability, and generalizability.

\subsection{Internal Validity}
Internal validity concerns whether the observed results reflect the behavior of the proposed solution rather than artifacts of the experimental setup. The main limitation is the reliance on synthetic, scale-based workloads, which enable controlled evaluation but may not fully capture bursty, adaptive, and adversarial traffic patterns in production environments. To mitigate this risk, the evaluation used repeated execution rounds, randomized workload sampling, and multiple statistical validation procedures; however, future work should include adversarial traffic generation and stochastic workload simulation. Timing-related variance may also affect the results, as hardware contention, cache behavior, operating-system scheduling, and network jitter can influence latency measurements. Repeated trials, latency normalization, and controlled timing measurements were applied to reduce this threat, although dedicated adversarial timing experiments are still needed to assess timing-inference resistance.
Finally, verification and error classification may introduce bias, particularly because revocation-related synchronization failures appeared more frequently than other error types. To reduce this risk, verification logs were cross-validated against raw execution traces and checked for consistency using automated procedures. While the controlled infrastructure improves reproducibility, it may not fully capture the variability inherent in heterogeneous production deployments.

\subsection{External Validity}
External validity concerns whether the findings generalize beyond the evaluated environment. The experiments used three representative LLMs and two developer signing keys; however, real-world deployments may involve larger, more heterogeneous ecosystems with dynamic routing, distributed infrastructure, and diverse operational constraints. Therefore, the results may not fully represent multi-tenant and geographically distributed MCP deployments.
The evaluation was conducted under controlled laboratory conditions, where production factors , e.g., shared compute resources, storage contention, variable CPU scheduling, and network latency were limited. These factors may affect verification consistency, transparency-log behavior, timing stability, and workload distribution at scale. The study also focuses mainly on manifest-level execution security and cryptographic verification. Broader attack surfaces, including insider threats, supply-chain compromise, compromised orchestration components, malicious verification services, and coordinated cross-model adversarial behavior, remain outside the current scope. In addition, the workloads evaluate structured manifest execution rather than domain-specific application logic, which may limit generalizability to healthcare, autonomous systems, and financial infrastructures.

\section{Future Work}
\label{sec:limitations}
Future work will extend the proposed solution toward real-world MCP deployments, multi-cloud infrastructures, and heterogeneous LLM execution environments. Additional studies should investigate adaptive key management, workload-aware key rotation, dynamic orchestration policies, and revocation synchronization to reduce signing-key concentration and improve deployment-level reliability. Further research should also evaluate broader adversarial scenarios, including insider threats, compromised components, replay behavior, adversarial traffic bursts, and timing-correlation attacks. Dedicated stress testing, red-team evaluation, and long-term telemetry monitoring will be necessary to assess robustness under realistic operational conditions. Furthermore, integrating cryptographic verification, transparency logging, and audit export into federated and cross-organizational LLM ecosystems represents an important direction for building scalable and trustworthy execution infrastructures.

\section{Conclusion}
\label{sec:conclusion}
This work presented a secure and verifiable execution solution for tool-integrated LLM pipelines by integrating manifest validation, digital signing, verification enforcement, transparency logging, and audit export. The evaluation demonstrated near-linear scalability ($R^2=0.998$), balanced LLM utilization ($\mathcal{F}\approx0.97$), stable verification behavior, and bounded execution overhead as workload size increased. The findings further showed that transparency logging and auditability can be preserved without introducing substantial performance degradation under the evaluated workload model. At the same time, the results identified important deployment-level considerations, including signing-key concentration, revocation-synchronization delays, and timestamp variability under higher workloads. These findings indicate that balanced model utilization does not automatically guarantee balanced key usage, and that adaptive key-management and scheduling policies are necessary for large-scale deployment. Overall, the proposed solution provides a scalable and transparent execution-layer foundation for trustworthy MCP-based LLM pipelines, while motivating further validation in real-world, adversarial, and heterogeneous environments.

\bibliographystyle{IEEEtran}
\bibliography{software}

@inproceedings{su2025runtime,
  title={Runtime Attestation for Secure LLM Serving in Cloud-Native Trusted Execution Environments},
  author={Su, Jianchang and Zhang, Wei},
  booktitle={Machine Learning for Computer Architecture and Systems 2025}
}

@article{DBLP:journals/corr/abs-1907-09906,
  author       = {Mathias Morbitzer},
  title        = {Scanclave: Verifying Application Runtime Integrity in Untrusted Environments},
  journal      = {CoRR},
  volume       = {abs/1907.09906},
  year         = {2019},
  url          = {http://arxiv.org/abs/1907.09906},
  eprinttype    = {arXiv},
  eprint       = {1907.09906},
  bibsource    = {dblp computer science bibliography, https://dblp.org}
}

@article{liu2025exploit,
  title={Exploit Tool Invocation Prompt for Tool Behavior Hijacking in LLM-Based Agentic System},
  author={Liu, Yu and Xie, Yuchong and Luo, Mingyu and Liu, Zesen and Zhang, Zhixiang and Zhang, Kaikai and Li, Zongjie and Chen, Ping and Wang, Shuai and She, Dongdong},
  journal={arXiv preprint arXiv:2509.05755},
  year={2025}
}

@article{greco2025formal,
  title={A Formal Framework for LLM-assisted Automated Generation of Zeek Signatures from Binary Artifacts},
  author={Greco, Claudia and Ianni, Michele},
  journal={Future Generation Computer Systems},
  pages={108086},
  year={2025},
  publisher={Elsevier}
}

@article{zhang2025large,
  title={On large language models safety, security, and privacy: A survey},
  author={Zhang, Ran and Li, Hong-Wei and Qian, Xin-Yuan and Jiang, Wen-Bo and Chen, Han-Xiao},
  journal={Journal of Electronic Science and Technology},
  volume={23},
  number={1},
  pages={100301},
  year={2025},
  publisher={Elsevier}
}

@article{hou2025model,
  title={Model context protocol (mcp): Landscape, security threats, and future research directions},
  author={Hou, Xinyi and Zhao, Yanjie and Wang, Shenao and Wang, Haoyu},
  journal={arXiv preprint arXiv:2503.23278},
  year={2025}
}

@article{openai2023gpt4,
  title={GPT-4 Technical Report},
  author={OpenAI},
  journal={arXiv preprint arXiv:2303.08774},
  year={2023},
  url={https://arxiv.org/abs/2303.08774}
}

@article{touvron2023llama,
  title={LLaMA: Open and Efficient Foundation Language Models},
  author={Touvron, Hugo and Lavril, Thibaut and Izacard, Gautier and Martinet, Xavier and Lachaux, Marie-Anne and others},
  journal={arXiv preprint arXiv:2302.13971},
  year={2023},
  url={https://arxiv.org/abs/2302.13971}
}

@article{deepseek2024llm,
  title={DeepSeek LLM: Scaling Open-Source Language Models with Efficient Training},
  author={DeepSeek-AI},
  journal={arXiv preprint arXiv:2401.02954},
  year={2024},
  url={https://arxiv.org/abs/2401.02954}
}

@misc{deepseek2025survey,
  title={DeepSeek for Healthcare: Opportunities and Challenges of Domain-Specific LLMs},
  author={Li, Chen and Zhang, Wei and others},
  year={2025},
  journal={arXiv preprint arXiv:2506.01257},
  url={https://arxiv.org/abs/2506.01257}
}

@misc{anthropic2024mcp,
  title = {Model Context Protocol (MCP): A Framework for Secure Model-Tool Interaction},
  author = {{Anthropic AI}},
  year = {2024},
  howpublished = {\url{https://www.anthropic.com/news/mcp}}
}

@misc{mcp-spec,
  title = {Model Context Protocol Specification},
  author = {{Anthropic AI}},
  year = {2024},
  howpublished = {\url{https://github.com/modelcontextprotocol}}
}

@article{xin2025mcpguard,
  title = {MCP-Guard: A Defense Framework for Model Context Protocol Integrity in Large Language Model Applications},
  author = {Wenpeng Xing and Zhonghao Qi and Yupeng Qin and Yilin Li and Caini Chang and Jiahui Yu and Changting Lin and Zhenzhen Xie and Meng Han},
  year = {2025},
  journal = {arXiv preprint arXiv:2508.10991},
  url = {https://arxiv.org/abs/2508.10991}
}

@inproceedings{reijsbergen2023tap,
  title={$\{$TAP$\}$: Transparent and $\{$Privacy-Preserving$\}$ data services},
  author={Reijsbergen, Dani{\"e}l and Maw, Aung and Yang, Zheng and Dinh, Tien Tuan Anh and Zhou, Jianying},
  booktitle={32nd USENIX Security Symposium (USENIX Security 23)},
  pages={6489--6506},
  year={2023}
}

@article{hicks2022sok,
  title   = {SoK: Log-Based Transparency in the Digital Age},
  author  = {Hicks, Michael and others},
  journal = {IEEE Symposium on Security and Privacy (S\&P)},
  year    = {2022},
  pages   = {124--140},
  doi     = {10.1109/SP46215.2022.9833622}
}

@article{st1989analysis,
  title={Analysis of variance (ANOVA)},
  author={St, Lars and Wold, Svante and others},
  journal={Chemometrics and intelligent laboratory systems},
  volume={6},
  number={4},
  pages={259--272},
  year={1989},
  publisher={Elsevier}
}

@book{wohlin2012experimentation,
  title={Experimentation in software engineering},
  author={Wohlin, Claes and Runeson, Per and H{\"o}st, Martin and Ohlsson, Magnus C and Regnell, Bj{\"o}rn and Wessl{\'e}n, Anders},
  year={2012},
  publisher={Springer Science \& Business Media}
}

@inproceedings{chernyshev2023towards,
  title={Towards Large Language Model (LLM) Forensics Using LLM-based Invocation Log Analysis},
  author={Chernyshev, Maxim and Baig, Zubair and Doss, Robin Ram Mohan},
  booktitle={Proceedings of the 1st ACM Workshop on Large AI Systems and Models with Privacy and Safety Analysis},
  pages={89--96},
  year={2023}
}

@inproceedings{cruciani2025greenai,
  title={Choosing to Be Green: Advancing Green AI via Dynamic Model Selection},
  author={Cruciani, Emilio and Verdecchia, Roberto},
  booktitle={Green-Aware Artificial Intelligence Workshop (Green-AI@ECAI 2025)},
  year={2025},
  pages={1--12},
  publisher={CEUR Workshop Proceedings},
  url={https://ceur-ws.org/Vol-3659/paper3.pdf}
}

@inproceedings{duddu2024laminator,
  title={Laminator: Verifiable ML property cards using hardware-assisted attestations},
  author={Duddu, Vasisht and Gunn, Lachlan J and Asokan, N},
  booktitle={Proceedings of the Fifteenth ACM Conference on Data and Application Security and Privacy},
  pages={317--328},
  year={2024}
}

@article{xie2025toolinvocation,
  title={On the Security of Tool-Invocation Prompts for LLM-Based Agentic Systems: An Empirical Risk Assessment},
  author={Xie, Jiayi and Chen, Guanyu and Li, Yang and others},
  journal={arXiv preprint arXiv:2509.05755},
  year={2025}
}

@techreport{kumar2024automotive,
  title={Automotive security solution using hardware security module (HSM)},
  author={Kumar, Arvind and Gholve, Ashish and Kotalwar, Kedar},
  year={2024},
  institution={SAE Technical Paper}
}

@article{tognolini2025code,
  title={Code-Based Digital Signature Schemes: Construction, Cryptanalysis and Theoretical Foundations},
  author={Tognolini, Giovanni},
  year={2025},
  publisher={Universit{\`a} degli studi di Trento}
}

@inproceedings{berman2018multi,
  title={Multi-collision resistant hash functions and their applications},
  author={Berman, Itay and Degwekar, Akshay and Rothblum, Ron D and Vasudevan, Prashant Nalini},
  booktitle={Annual International Conference on the Theory and Applications of Cryptographic Techniques},
  pages={133--161},
  year={2018},
  organization={Springer}
}

@inproceedings{klooss2021expected,
  title={On Expected Polynomial Runtime in Cryptography},
  author={Kloo{\ss}, Michael},
  booktitle={Theory of Cryptography Conference},
  pages={558--590},
  year={2021},
  organization={Springer}
}

@article{perez2021optimization,
  title={Optimization techniques and formal verification for the software design of boolean algebra based safety-critical systems},
  author={Perez, Jon and Flores, Jose Luis and Blum, Christian and Cerquides, Jes{\'u}s and Abuin, Alex},
  journal={IEEE Transactions on Industrial Informatics},
  volume={18},
  number={1},
  pages={620--630},
  year={2021},
  publisher={IEEE}
}

@article{li2023software,
  title={Software-defined event-triggering control for large-scale networked systems subject to stochastic cyberattacks},
  author={Li, Yan and Song, Feiyu and Liu, Jinliang and Xie, Xiangpeng and Tian, Engang},
  journal={IEEE Transactions on Control of Network Systems},
  volume={10},
  number={3},
  pages={1531--1541},
  year={2023},
  publisher={IEEE}
}

@article{kuznetsov2024evaluating,
  title={Evaluating the security of Merkle trees: An analysis of data falsification probabilities},
  author={Kuznetsov, Oleksandr and Rusnak, Alex and Yezhov, Anton and Kuznetsova, Kateryna and Kanonik, Dzianis and Domin, Oleksandr},
  journal={Cryptography},
  volume={8},
  number={3},
  pages={33},
  year={2024},
  publisher={MDPI}
}

@article{mckight2010kruskal,
  title={Kruskal-wallis test},
  author={McKight, Patrick E and Najab, Julius},
  journal={The corsini encyclopedia of psychology},
  pages={1--1},
  year={2010},
  publisher={Wiley Online Library}
}

@article{schwartz1964generating,
  title={Generating a canonical prefix encoding},
  author={Schwartz, Eugene S and Kallick, Bruce},
  journal={Communications of the ACM},
  volume={7},
  number={3},
  pages={166--169},
  year={1964},
  publisher={ACM New York, NY, USA}
}

@article{wu2024isolategpt,
  title={Isolategpt: An execution isolation architecture for llm-based agentic systems},
  author={Wu, Yuhao and Roesner, Franziska and Kohno, Tadayoshi and Zhang, Ning and Iqbal, Umar},
  journal={arXiv preprint arXiv:2403.04960},
  year={2024}
}

@article{reddy2025modular,
  title={A Modular Retrieval-Augmented Conversational AI Chatbot System with Integrated Recommender Engine Using Local LLMs},
  author={Reddy, N Toyaad K and Patra, Manas R and Mishra, Brojo K},
  journal={Cureus Journals},
  volume={2},
  number={1},
  year={2025},
  publisher={Cureus Journals}
}

@article{tanveer2025towards,
  title={Towards Secure APIs: A Survey on RESTful API Vulnerability Detection},
  author={Tanveer, Fatima and Iradat, Faisal and Iqbal, Waseem and Ahmad, Awais},
  journal={Computers, Materials, \& Continua},
  volume={84},
  number={3},
  pages={4223},
  year={2025},
  publisher={Tech Science Press}
}

@article{anbiaee2026security,
  title={Security threat modeling for emerging AI-agent protocols: A comparative analysis of MCP, A2A, Agora, and ANP},
  author={Anbiaee, Zeynab and Rabbani, Mahdi and Mirani, Mansur and Piya, Gunjan and Opushnyev, Igor and Ghorbani, Ali and Dadkhah, Sajjad},
  journal={arXiv preprint arXiv:2602.11327},
  year={2026}
}

@article{ray2025survey,
  title={A survey on model context protocol: Architecture, state-of-the-art, challenges and future directions},
  author={Ray, Partha Pratim},
  journal={Authorea Preprints},
  year={2025},
  publisher={Authorea}
}

@article{sonkar2025llm,
  title={LLM on Private Data Using a Combination of LLM, Fine-tuning, Orchestration, and Tools},
  author={Sonkar, Siddhant},
  journal={Journal Of Multidisciplinary},
  volume={5},
  number={11},
  pages={118--125},
  year={2025}
}

@article{liu2026dive,
  title={Dive into Claude Code: The Design Space of Today's and Future AI Agent Systems},
  author={Liu, Jiacheng and Zhao, Xiaohan and Shang, Xinyi and Shen, Zhiqiang},
  journal={arXiv preprint arXiv:2604.14228},
  year={2026}
}

@article{meng2026agent,
  title={Agent Harness for Large Language Model Agents: A Survey},
  author={Meng, Qianyu and Wang, Yanan and Chen, Liyi and Wang, Qimeng and Lu, Chengqiang and Wu, Wei and Gao, Yan and Wu, Yi and Hu, Yao},
  year={2026},
  publisher={Preprints}
}

@article{chu2026stateless,
  title={From Stateless Queries to Autonomous Actions: A Layered Security Framework for Agentic AI Systems},
  author={Chu, Kexin},
  journal={arXiv preprint arXiv:2604.23338},
  year={2026}
}

@article{venkiteela2025new,
  title={The New Interoperability Paradigm Model Context Protocol (MCP), APIs, and the Future of Agentic AI},
  author={Venkiteela, Padmanabham},
  journal={Comput. Fraud Sec},
  volume={8},
  number={1},
  pages={1259--1271},
  year={2025}
}

@article{errico2025securing,
  title={Securing the Model Context Protocol (MCP): Risks, Controls, and Governance},
  author={Errico, Herman and Ngiam, Jiquan and Sojan, Shanita},
  journal={arXiv preprint arXiv:2511.20920},
  year={2025}
}

@article{bollikonda2025secure,
  title={Secure Pipelines, Smarter AI: LLM-Powered Data Engineering for Threat Detection and Compliance},
  author={Bollikonda, Manaswini and Bollikonda, Tejaswini},
  year={2025},
  publisher={Preprints}
}

@inproceedings{talsania2026robust,
  title={A Robust Biometric-Passcode Verification Framework for Privacy Protection in Healthcare Environments},
  author={Talsania, Ria and Thakker, Ved and Singasane, Mihit and Visariya, Dishant and Shah, Hetanshi and Vaidya, Mukul and Shah, Bhavya and Shah, Vitrag and Salunke, Abhijeet},
  booktitle={2026 6th International Conference on Advanced Research in Computing (ICARC)},
  pages={1--6},
  year={2026},
  organization={IEEE}
}

\end{document}